\documentclass[letterpaper,english,letterpaper,english,aps, twocolumn, prb, longbibliography]{revtex4-1}
\usepackage[T1]{fontenc}
\usepackage[latin9]{inputenc}
\usepackage{babel}
\usepackage{bm}
\usepackage{amsmath}
\usepackage{amssymb}
\usepackage{graphicx}
\usepackage[unicode=true,pdfusetitle,
 bookmarks=true,bookmarksnumbered=false,bookmarksopen=false,
 breaklinks=false,pdfborder={0 0 1},backref=false,colorlinks=false]
 {hyperref}



\begin{document}
\title{Effect of boundary roughness on the attenuation of specular phonon
reflection in graphene}
\author{Zhun-Yong Ong}
\affiliation{Institute of High Performance Computing (IHPC), Agency for Science,
Technology and Research (A{*}STAR), 1 Fusionopolis Way, \#16-16 Connexis,
Singapore 138632, Republic of Singapore}
\email{ongzy@ihpc.a-star.edu.sg}

\date{\today}
\begin{abstract}
The reduced phonon specularity $p$ from boundary roughness scattering
plays a major role in the lower thermal conductivity in semiconducting
and insulating nanowires and films. Although the well-known Ziman
formula $p=\exp(-4\sigma^{2}q_{x}^{2})$, where $\sigma$ and $q_{x}$
denote the root-mean-square boundary roughness and the normal component
of the incident phonon wave vector, respectively, and its variants
are commonly used in the literature to estimate how roughness attenuates
$p$, their validity and accuracy remain poorly understood, especially
when the effects of mode conversion cannot be ignored. In this paper,
we investigate the accuracy and validity of the more general Ogilvy
formula, from which the Ziman formula is derived, by comparing its
predictions to the $p$ values computed from Atomistic Green's Function
(AGF) simulations for an ensemble of rough boundaries in single-layer
graphene. The effects of phonon dispersion, incident angle, polarization,
mode conversion, and correlation length are analyzed. Our results
suggest that the Ogilvy formula is remarkably accurate for $0<q_{x}<\frac{\pi}{4\sigma}$
when the lateral correlation length $L$ is large or the phonon is
at normal incidence. At large $q_{x}$ in the short-wavelength limit,
the $q_{x}$-dependence of $p$ becomes significantly weaker. In the
large-$L$ limit, the numerical results suggest the existence of a
minimum $p$ for short-wavelength phonons, given by $p\sim p_{0}\exp(-\pi^{2}/4)$,
where $p_{0}$ is the baseline specularity for the ideal boundary. 
\end{abstract}
\maketitle

\section{Introduction}

The scattering of waves by rough boundaries, or \emph{boundary roughness
scattering} for short, is an area of longstanding interest in a wide
range of fundamental and applied fields such as optics~\citep{SNee:AO96_Polarization},
computer graphics~\citep{BVanGinneken:ApplOpt98_Diffuse}, acoustics~\citep{MDarmon:AS20_Acoustic},
non-destructive testing~\citep{RDwyerJoyce:JTrib00_Use,SHaslinger:Insight21_Prediction},
seismology~\citep{JRobertsson:GJI06_Modelling}, and nanoscale thermal
transport~\citep{PMartin:PRL09_Impact,AMaznev:PRB15_Boundary,FShi:PRB17_Diffusely,NRavichandran:PRX18_Spectrally,DGelda:PRB18_Specularity}.
A problem of particular significance in boundary roughness scattering
is the attenuated intensity of the specularly reflected wave, as characterized
by the \emph{specular reflectance}, which refers to the proportion
of an incident wave that is scattered into the specular direction.
For an ideal flat boundary, the incident wave is completely reflected
in the specular direction because the \emph{continuous} translational
symmetry of the boundary requires the transverse component of the
incident and reflected wave vectors in the plane of incidence (denoted
by $\boldsymbol{q}_{i}$ and $\boldsymbol{q}_{r}$, respectively)
to be conserved. This conservation condition is described by the generalized
Snell's law, i.e., $q_{i}^{\parallel}=q_{r}^{\parallel}$, where the
$\parallel$ superscript denotes the transverse component of the wave
vector. For a nonideal boundary, the boundary roughness breaks this
translational symmetry, resulting in the phenomenon of attenuated
specular reflectance where the amplitude of the specularly reflected
wave is reduced as part of the reflected wave is scattered in the
non-specular directions.~\citep{FShi:JMPS21_Variance}

In condensed matter physics, the phenomenon of boundary roughness
scattering plays an important role in the reduced lattice thermal
conductivity and thermoelectric properties of low-dimensional semiconductors
(e.g. silicon nanowires and nanofilms). In these materials, thermal
conduction in the bulk is primarily mediated by the propagation of
phonons, the wavelike crystal lattice excitations that undergo scattering
with the lattice boundary. In materials with rough boundaries~\citep{PMartin:PRL09_Impact,JLim:NL12_Quantifying,DGelda:PRB18_Specularity,NRavichandran:PRX18_Spectrally},
this boundary roughness scattering decreases the specular reflectance
of the phonons and results in phonon momentum dissipation. For an
incident wave or phonon, the degree of specular reflectance is characterized
by the specularity parameter $p$, which represents the probability
of specular reflection. When there is perfect specular reflection
by an ideal boundary, we have $p=1$ and no momentum dissipation in
the axial direction of propagation. In the diffuse limit for an infinitely
rough boundary, the incident phonon is assumed to be scattered equally
in all directions such that $p=0$. For a nonideal boundary of finite
roughness, we expect $p$ to vary between 0 and 1, with the numerical
value depending on the degree of boundary roughness, and this results
in phonon momentum dissipation and resistance to thermal transport.
In addition, the effective momentum exchange between the phonon and
the boundary, which determines the impact of scattering on the thermal
conductivity, also depends on the form of the diffuse field and the
power spectrum of the boundary structure~\citep{JLim:NL12_Quantifying}. 

In spite of its relevance for understanding lattice thermal conduction,
the boundary roughness scattering of phonons remains poorly characterized,
especially in terms of the dependence of $p$ on the properties of
the incident phonon (incident angle, polarization and crystal momentum)
and the structure of the boundary. For the benefit of the reader,
we give a brief overview of the current theoretical description of
the phenomenon. Current theories of how boundary roughness scattering
attenuates $p$ rely heavily on analogies of scalar wave scattering
by rough boundaries. In the simplest model, $p$ is related to the
boundary roughness through the \emph{Rayleigh formula}~\citep{JAOgilvy:RPR87_Wave,JDeSanto:Book92_Scalar,ZAksamija:APL11_Lattice}
\begin{equation}
p=p_{0}\exp(-4\Sigma^{2})\ ,\label{eq:RayleighFormula}
\end{equation}
where $\Sigma$ and $p_{0}$ denote the so-called Rayleigh roughness
parameter~\citep{JDeSanto:Book92_Scalar} and the specularity parameter
for the ideal boundary, respectively. In Eq.~(\ref{eq:RayleighFormula}),
we define $\Sigma$ as $\Sigma=\sigma q\cos\theta_{i}$, where $q$,
$\sigma$, and $\theta_{i}$ denote the incident wave number, the
root-mean-square (RMS) boundary roughness, and the angle of incidence,
respectively. It should be noted that $p$ and $p_{0}$ are functions
of $\theta_{i}$ and $q=\frac{2\pi}{\lambda}$ where $\lambda$ is
the incident wavelength. Physically, Eq.~(\ref{eq:RayleighFormula})
describes the exponential attenuation of the specularity parameter. 

To be applicable, Eq.~(\ref{eq:RayleighFormula}) must satisfy the
Rayleigh roughness criterion, i.e., $\Sigma<\pi/4$~\citep{JDeSanto:Book92_Scalar},
which relates the boundary roughness to the incident wavelength, because
the attenuation factor of $\exp(-4\Sigma^{2})$ originates from the
phase interference between random vertically aligned sites on the
boundary. At larger values of $\sigma$ or shorter incident wavelengths
(larger $q$'s), Eq.~(\ref{eq:RayleighFormula}) is not expected
to be valid when $q>\frac{\pi}{4\sigma}$ or $\lambda<8\sigma$ although
there is some uncertainty over the degree of its discrepancy with
the actual $p$. In addition, Eq.~(\ref{eq:RayleighFormula}) does
not give an explicit dependence on the degree of undulation of the
boundary, as characterized by its autocorrelation function $\langle h(\bm{r})h(0)\rangle$,
where $h(\bm{r})$ is the displacement of the boundary from its mean
position at the point $\bm{r}$, although Eq.~(\ref{eq:RayleighFormula})
is derived from the \emph{Kirchhoff Approximation} which assumes that
the lateral correlation length $L$ is much greater than the wavelength,
i.e., $qL\gg1$~\citep{AMaznev:PRB15_Boundary}. This means that
Eq.~(\ref{eq:RayleighFormula}) is also not valid at long wavelengths
when $q\ll\frac{1}{L}$. 

Taken together, the Rayleigh roughness criterion and the Kirchhoff
Approximation imply that the boundary roughness scattering for only
a finite range of incident waves, as band-limited by the condition
$\frac{1}{L}\ll q\lesssim\frac{\pi}{4\sigma}$, can be described by
Eq.~(\ref{eq:RayleighFormula}). A corollary of this is that Eq.~(\ref{eq:RayleighFormula})
is not expected to be valid if $L\lesssim\sigma$. Equation~(\ref{eq:RayleighFormula})
cannot describe the boundary scattering of long-wavelength waves unlesss
the lateral correlation length is larger than the wavelength and it
also cannot describe the boundary scattering of short-wavelength waves
unless the RMS boundary roughness is smaller than the wavelength.
This is succinctly described by the condition 
\begin{equation}
\sigma<\lambda\ll L\ .\label{eq:RayleighValidityCondition}
\end{equation}
In addition, it is common in the literature to ignore the angular
dependence in Eq.~(\ref{eq:RayleighFormula}) and simply set $p=\exp(-4\sigma^{2}q^{2})$,
an expression that is sometimes attributed to Ziman~\citep{DGelda:PRB18_Specularity,NRavichandran:PRX18_Spectrally}
and significantly overestimates the phonon momentum dissipation from
boundary roughness scattering especially for phonons impinging on
the boundary at a grazing angle~\citep{AMaznev:PRB15_Boundary}.

In a crystal lattice, the wavelike phonons can similarly undergo scattering
by boundary roughness although two factors constrain the applicability
of Eq.~(\ref{eq:RayleighFormula}) for understanding how boundary
roughness attenuates specular reflection. The first is the discrete
lattice structure of the solid which sets a minimum length scale absent
from Eq.~(\ref{eq:RayleighFormula}) and eliminates the continuous
translational symmetry assumed in scalar wave models. The second is
scattering-induced \emph{mode conversion} which changes the polarization
of the incoming phonon. For example, an incoming longitudinal acoustic
(LA) phonon can be scattered and transformed into an outgoing transverse
acoustic (TA) phonon with a finite probability that depends on the
angle of incidence. The additional effect of mode conversion means
that the attenuation by boundary roughness is not a simple interference
effect, as in the case of scalar wave scattering~\citep{JDeSanto:Book92_Scalar},
but must account for the vectorial nature of the atomic displacement. 

\subsection{The Ogilvy formula for phonons}

Nevertheless, Eq.~(\ref{eq:RayleighFormula}) lends itself to a possible
generalization~\citep{JAOgilvy:RPR87_Wave} that takes mode conversion
into account. Let us associate each phonon with a polarization $\nu$
and wave vector $\boldsymbol{q}$ and use the subscripts $i$ and
$r$ to label the incident and reflected phonons, respectively. We
use the \emph{mode-dependent} function $p_{\sigma}(\nu_{r}\boldsymbol{q}_{r},\nu_{i}\boldsymbol{q}_{i})$
to denote the transition probability that an incident $\nu_{i}\boldsymbol{q}_{i}$
phonon is \emph{specularly} reflected into an outgoing $\nu_{r}\boldsymbol{q}_{r}$
phonon by a boundary of RMS roughness $\sigma$, and we can thus interpret
$p_{\sigma}(\nu_{r}\boldsymbol{q}_{r},\nu_{i}\boldsymbol{q}_{i})$
intuitively as the ratio of the probability \emph{flux} of the outgoing
and incoming phonon modes given by $I^{\text{out}}(\nu_{r}\boldsymbol{q}_{r})$
and $I^{\text{in}}(\nu_{i}\boldsymbol{q}_{i})$, respectively. In
other words, $p_{\sigma}(\nu_{r}\boldsymbol{q}_{r},\nu_{i}\boldsymbol{q}_{i})=I^{\text{out}}(\nu_{r}\boldsymbol{q}_{r})/I^{\text{in}}(\nu_{i}\boldsymbol{q}_{i})$.
Physically, $I^{\text{in}}(\nu_{i}\boldsymbol{q}_{i})$ describes
the \emph{incoming} flux of phonons with polarization $\nu_{i}$ (e.g.
longitudinal acoustic) and wave vector $\bm{q}_{i}$. Likewise, $I^{\text{out}}(\nu_{r}\boldsymbol{q}_{r})$
describes the \emph{outgoing} flux of phonons with polarization $\nu_{r}$
and wave vector $\bm{q}_{r}$. The wave vectors $\boldsymbol{q}_{i}$
and $\boldsymbol{q}_{r}$ are not independent variables but are related
through the generalized Snell's law ($q_{i}^{\parallel}=q_{r}^{\parallel}$)
that expresses the conservation of transverse momentum. 

Therefore, a natural generalization of Eq.~(\ref{eq:RayleighFormula})
leads us to the expression~\citep{AMaznev:PRB15_Boundary} 
\begin{equation}
p_{\sigma}(\nu_{r}\boldsymbol{q}_{r},\nu_{i}\boldsymbol{q}_{i})=p_{0}(\nu_{r}\boldsymbol{q}_{r},\nu_{i}\boldsymbol{q}_{i})\exp[-\sigma^{2}(|q_{r}^{\perp}|+|q_{i}^{\perp}|)^{2}]\ ,\label{eq:ProbOgilvyFormula}
\end{equation}
where $p_{0}(\nu_{r}\boldsymbol{q}_{r},\nu_{i}\boldsymbol{q}_{i})=p_{\sigma}(\nu_{r}\boldsymbol{q}_{r},\nu_{i}\boldsymbol{q}_{i})|_{\sigma=0}$
denotes the specularity parameter for the ideal boundary and the $\perp$
superscript on the right-hand side of Eq.~(\ref{eq:ProbOgilvyFormula})
denotes the longitudinal component of the wave vector that is at normal
incidence to the boundary. If there is no mode conversion ($\nu_{r}=\nu_{i}$),
then $|q_{r}^{\perp}|=|q_{i}^{\perp}|$ and we recover the Rayleigh
formula in Eq.~(\ref{eq:RayleighFormula}). For the attenuation of
elastodynamic waves by boundary roughness scattering, an analogous
expression is given by Ogilvy in Ref.\ \citep{JAOgilvy:RPR87_Wave},
which we refer to as the \emph{Ogilvy formula} for convenience in
the rest of this article. Because phonons in the long wavelength limit
are described by elastodynamic waves, we expect the Ogilvy formula
to be applicable for the boundary roughness of phonons even though
its validity and accuracy remain untested. 

Nonetheless, our understanding of the accuracy of Eq.~(\ref{eq:ProbOgilvyFormula})
for estimating the phonon specularity is poor, partly because of the
challenges in the experimental measurement of individual phonon amplitudes~\citep{GNorthrop:PRL84_Phonon,DGelda:PRB18_Specularity,NRavichandran:PRX18_Spectrally}.
Instead, the simpler Eq.~(\ref{eq:RayleighFormula}), which ignores
the effects of polarization, is more commonly used in the literature
especially for the interpretation of experimentally measured thermal
conductivity values~\citep{NRavichandran:PRX18_Spectrally}. Even
with the use of simulations~\citep{CShao:JAP17_Probing}, it is difficult
to assess the accuracy of Eq.~(\ref{eq:ProbOgilvyFormula}) because
of the computational difficulties in isolating the specularly reflected
wave polarization components after the scattering. Furthermore, it
is unclear how the nonlinear dispersion and discrete symmetries for
phonons affect the validity of the Ogilvy formula which is derived
for elastodynamic waves in a continuum solid~\citep{JAOgilvy:RPR87_Wave}. 

It is worth noting, however, that the results from an experimental
study of the boundary scattering of phonons in silicon nanosheets~\citep{JHertzberg:NL14_Direct}
cast some doubt on the validity of Eqs.~(\ref{eq:RayleighFormula})
and (\ref{eq:ProbOgilvyFormula}). In Ref.~\citep{JHertzberg:NL14_Direct},
the measured phonon specularity values are much smaller than those
predicted using the Ziman theory, which is based on Eq.~(\ref{eq:RayleighFormula}),
with the geometrical boundary roughness of the silicon nanosheets
used as the input. One possible explanation for the discrepancy is
that Eq.~(\ref{eq:RayleighFormula}) is not applicable to nanosheets,
where the system has two opposite rough surfaces. Another is that
Eqs.~(\ref{eq:RayleighFormula}) and (\ref{eq:ProbOgilvyFormula})
are applicable but with the \emph{effective} boundary roughness associated
with scattering being much greater than the geometrical boundary roughness
due to the changes in the morphology of the surfaces from amorphization
or oxidation.

\subsection{Direct calculation of probability of specular reflection}

However, the opportunity for assessing the accuracy of Eq.~(\ref{eq:ProbOgilvyFormula})
has been greatly improved by extensions of the Atomistic Green's Function
(AGF) method~\citep{NMingo:PRB03_Phonon} for modeling mode-resolved
phonon transmission and reflection~\citep{ZYOng:JAP18_Tutorial,ZYOng:PRB18_Atomistic}
and recent advances in the identification of phonon polarization in
lattice models~\citep{CGan:JPC21_Complementary}. These improved
computational techniques allow us to efficiently identify the outgoing
$\nu_{r}\boldsymbol{q}_{r}$ and the incoming $\nu_{i}\boldsymbol{q}_{i}$
phonon modes and hence calculate their scattering amplitude $S_{\sigma}(\nu_{r}\boldsymbol{q}_{r},\nu_{i}\boldsymbol{q}_{i})$,
as defined in the equation 
\begin{equation}
\psi^{\text{out}}(\nu_{r}\boldsymbol{q}_{r})=S_{\sigma}(\nu_{r}\boldsymbol{q}_{r},\nu_{i}\boldsymbol{q}_{i})\psi^{\text{inc}}(\nu_{i}\boldsymbol{q}_{i})\label{eq:ScatterAmplitude}
\end{equation}
where $\psi^{\text{out}}$ and $\psi^{\text{inc}}$ are the \emph{complex
flux amplitudes} for the outgoing and incoming phonon modes, respectively,
for the given lattice realization with the boundary roughness $\sigma$.
Given that $S_{\sigma}(\nu_{r}\boldsymbol{q}_{r},\nu_{i}\boldsymbol{q}_{i})$
is a random variable, only ensemble averages of variables are meaningful.
Thus, from Eq.~(\ref{eq:ScatterAmplitude}), we obtain the exact
expression for the probability of specular reflection, i.e., 
\begin{equation}
p_{\sigma}(\nu_{r}\boldsymbol{q}_{r},\nu_{i}\boldsymbol{q}_{i})=\langle|S_{\sigma}(\nu_{r}\boldsymbol{q}_{r},\nu_{i}\boldsymbol{q}_{i})_{q_{r}^{\parallel}=q_{i}^{\parallel}}|^{2}\rangle\ ,\label{eq:ProbSpecReflection}
\end{equation}
where $\langle\ldots\rangle$ denotes the ensemble average of configurations
with the same boundary configuration (roughness and correlation length),
since $p_{\sigma}(\nu_{r}\boldsymbol{q}_{r},\nu_{i}\boldsymbol{q}_{i})=\langle I^{\text{out}}(\nu_{r}\boldsymbol{q}_{r})/I^{\text{in}}(\nu_{i}\boldsymbol{q}_{i})\rangle$
with $I^{\text{out}}(\nu_{r}\boldsymbol{q}_{r})=|\psi^{\text{out}}(\nu_{r}\boldsymbol{q}_{r})|^{2}$
and $I^{\text{in}}(\nu_{r}\boldsymbol{q}_{r})=|\psi^{\text{in}}(\nu_{r}\boldsymbol{q}_{r})|^{2}$.
The subscript $q_{r}^{\parallel}=q_{i}^{\parallel}$ for $S_{\sigma}$
on the righthand side of Eq.~(\ref{eq:ProbSpecReflection}) indicates
the conservation of transverse momentum in the scattering process.
Therefore, using Eq.~(\ref{eq:ProbSpecReflection}), we can test
the validity of Eq.~(\ref{eq:ProbOgilvyFormula}) because we can
compute $p_{\sigma}$ and $p_{0}$ directly from the scattering amplitudes
for a rough and flat boundary, respectively, with the AGF method.

\subsection{Scope and organization of the paper}

In this paper, the aim and scope of our investigation are quite modest
and specific. The primary object of our investigation is the extent
of the validity of the Ogilvy formula from Eq.~(\ref{eq:ProbOgilvyFormula}),
which was originally derived for elastodynamic waves~\citep{JAOgilvy:RPR87_Wave},
for describing the \emph{exponential attenuation} of the mode-dependent
$p_{\sigma}$ from boundary roughness scattering in single-layer graphene
(SLG). This is accomplished by computing the probability of specular
reflection $p_{\sigma}$ in Eq.~(\ref{eq:ProbSpecReflection}) directly
with the extended AGF method.   We investigate how the attenuation
of $p_{\sigma}$ varies with respect to the properties of the boundary
(e.g. roughness and correlation length) as well as the properties
of the phonons (e.g. wave vector, polarization, and angle of incidence).
It is hoped that our paper will shed light on some of the issues raised
in earlier studies on boundary roughness scattering and provide complementary
insights into existing work on phonon-boundary scattering~\citep{AMaznev:PRB15_Boundary,FShi:PRB17_Diffusely,CShao:JAP17_Probing,CShao:PRB18_Understanding}.
By directly calculating $p_{\sigma}$ for a range of incident wave
vectors $\bm{q}_{i}$, we circumvent some of the constraints (e.g.
weak roughness, long wavelengths, linear dispersion and normal incidence)
that arise from the approximations (e.g. the Kirchhoff approximation
and the small-perturbation method) used in other work~\citep{AMaznev:PRB15_Boundary,FShi:PRB17_Diffusely}.
This allows us to probe the effects of boundary roughness scattering
on $p_{\sigma}$ at grazing angles and short wavelengths comparable
to or smaller than the RMS roughness. In particular, we address two
issues that are not accessible in other methods or approximation but
can be treated with the extended AGF method. The first one pertains
to what happens to the attenuation of $p_{\sigma}$ at very small
wavelengths. The second one is on how the attenuation differs when
mode conversion takes place.

As our model of boundary roughness scattering in a condensed matter
system, we use a semi-infinite SLG lattice terminated by a stress-free
boundary with in-plane roughness. We choose SLG as our model system
for the following reasons. The first is that its two-dimensional (2D)
lattice reduces the computational load of calculating the scattering
amplitude as there is only one transverse dimension. The second is
that its out-of-plane flexural acoustic (ZA) phonons, which have a
quadratic dispersion ($\omega\propto k^{2}$) in the long wavelength
limit, allow us to study the effect of a nonlinear dispersion on Eq.~(\ref{eq:ProbOgilvyFormula}).
Also, because the boundary roughness is in-plane, the preservation
of the symmetry in the out-of-plane direction means that there is
no mode conversion between the flexural phonons and the in-plane phonons,
simplifying our analysis of the applicability of Eq.~(\ref{eq:ProbOgilvyFormula})
as mode conversion can only occur between the in-plane longitudinal
acoustic (LA) and transverse acoustic (TA) phonons. Thirdly, graphene
has been promoted as a material for heat spreading in the thermal
management of devices~\citep{YKuang:IJHMT16_Thermal} because of
its high native thermal conductivity. Thus, a clearer picture of boundary
roughness scattering in graphene is highly desirable for understanding
its thermal conductivity.

The organization of our paper is as follows. We discuss the statistical
description of the boundary roughness in graphene which we characterize
using the topographic parameter $\mathcal{T}$. The method for generating
the graphene boundary structure for a given RMS roughness $\sigma$
and lateral correlation length $L$ is described. We give an overview
of the properties of the \emph{bulk} acoustic phonons in graphene
because the phonons are the wavelike excitations that undergo scattering.
We then describe in some detail the AGF-based $S$-matrix method used
in computing $p_{\sigma}$. For convenience in the characterization
of the attenuation of $p_{\sigma}$, we introduce the mode-dependent
\emph{attenuation parameter} $\chi$ which is related to the Rayleigh
roughness parameter $\Sigma$. The dependence of $p_{\sigma}$ and
$\chi$, which we compute from our simulation results, on the boundary
structure and bulk acoustic phonon properties is discussed. In particular,
we pay attention to the areas where $\chi$ agrees and disagrees with
the predictions of the Ogilvy formula, and we characterize the behavior
of $\chi$ when the Ogilvy formula is not expected to be valid (e.g.
short wavelengths). We also measure the effective boundary roughness
$\rho_{\text{fit}}$, which we determine from fitting the attenuation
of $p_{\sigma}$, and we compare it to the geometrical RMS roughness
$\sigma$ of the structure. 

\section{Graphene boundary scattering}

\subsection{Atomistic model of rough graphene boundary}

\subsubsection{Statistical characterization of boundary roughness}

In our setup, the semi-infinite SLG sheet, which extends indefinitely
to the left ($x<0$), is located on the $x$-$y$ plane and terminated
by a boundary with its mean position ($x=0$) parallel to the $y$
axis. Thus, the longitudinal and transverse directions are parallel
to the $x$ and $y$ axis, respectively. Along the $x=0$ line, the
edge of the rough boundary is statistically characterized by the \emph{continuous}
random displacement function $h(y)$ which satisfies the Gaussian
correlation function~\citep{AMaznev:PRB15_Boundary} 
\begin{equation}
\langle h(y)h(0)\rangle=\sigma^{2}\exp(-y^{2}/L^{2})\ ,\label{eq:GaussianRoughness}
\end{equation}
where $\sigma=\sqrt{\langle h\rangle^{2}}$ and $L$ denote the RMS
roughness and the lateral correlation length along the $y$ axis,
respectively. We also define the mean displacement $\overline{h(y)}$
to be zero. The boundary is characterized by two length-scale parameters:
$R_{0}=2\sqrt{3}a_{\text{cc}}$ for the RMS roughness and $L_{0}=3a_{\text{cc}}$
for the correlation length where $a_{\text{cc}}$ is the equilibrium
carbon-carbon (C-C) bond length in bulk graphene. The correlation
length $L$ along the $y$ axis can be interpreted as the characteristic
\emph{feature size} of the boundary. When $L$ is small (large), the
boundary has a greater (smaller) lineal density of peaks and valleys. 

To characterize the statistical topography or loosely speaking, the
degree of jaggedness of the boundary, we also define and use the dimensionless
parameter $\mathcal{T}\equiv\sigma/L$, which we will refer to as
the \emph{topographic parameter} in the rest of the article. When
$\mathcal{T}$ is large (small), the boundary is more (less) jagged.
In general for the boundary roughness scattering of a scalar wave,
the Rayleigh roughness criterion and the Kirchhoff Approximation taken
together imply that the Ogilvy formula in Eq.~(\ref{eq:ProbOgilvyFormula})
is only valid when $\mathcal{T}\lesssim1$ in addition to the condition
in Eq.~(\ref{eq:RayleighValidityCondition}). We hypothesize that
this is also true for the boundary roughness scattering of phonons
in a graphene lattice and that the accuracy of the Ogilvy formula
increases as $\mathcal{T}$ decreases (i.e. less jagged).

\subsubsection{Atomistic realization of the rough graphene boundary}

For a graphene lattice, the orientation can be classified as either
``zigzag-edge'' or ``armchair-edge'' to describe the translational
symmetry of the lattice in the $y$ direction. In a zigzag-edge graphene
lattice, the arrangement of the C-C bonds in the $y$ direction has
a zigzag-like pattern. Likewise, in an armchair-edge graphene lattice,
the arrangement is armchair-like. In our paper, we limit the scope
of our investigation to the boundaries of zigzag-edge graphene for
convenience as the periodicity of the zigzag-edge boundary in the
$y$ direction is smaller and thus closer to a smooth boundary. 

\begin{figure}
\includegraphics[scale=0.5]{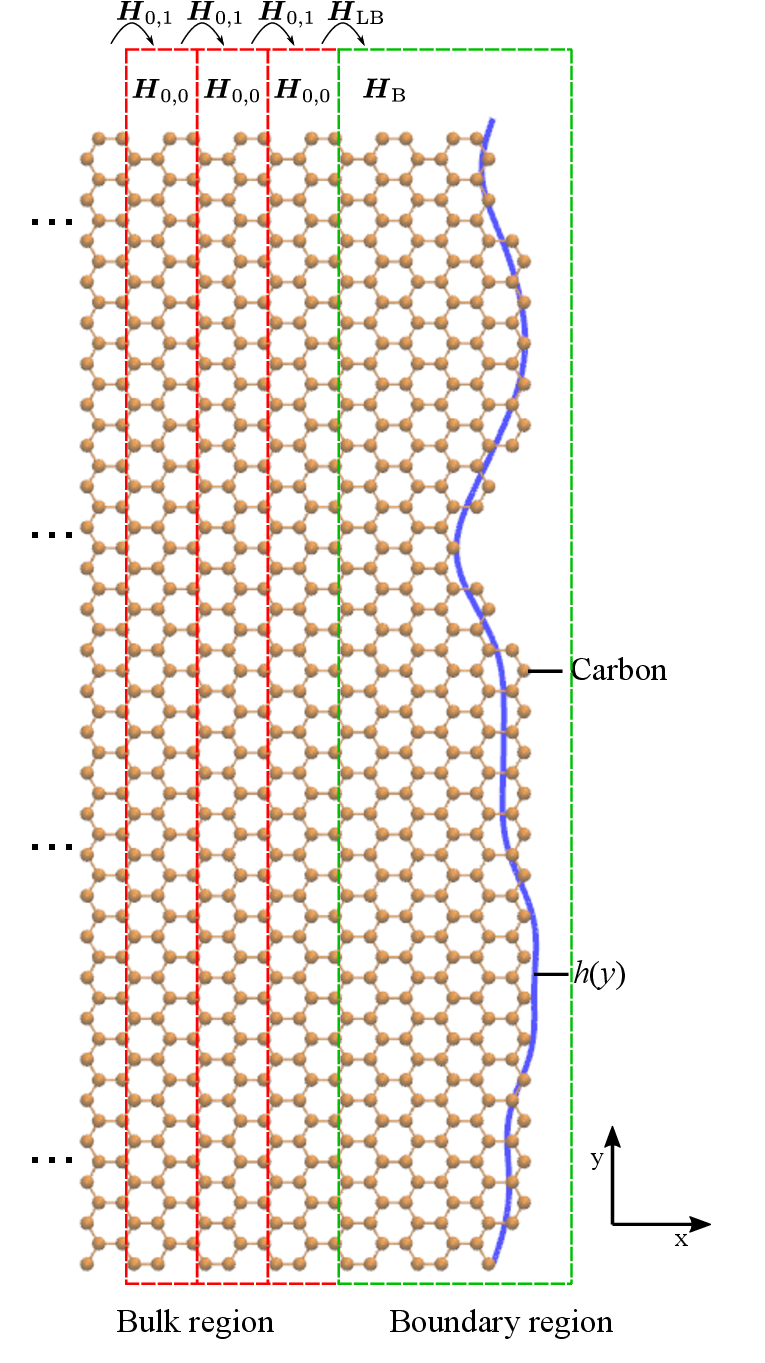} 

\caption{We plot of a realization of $h(y)$, indicated by the blue solid line,
and superimpose it on the corresponding lattice structure of the zigzag-edge
boundary of graphene for $\sigma=0.5R_{0}$ and $L=L_{0}$. We impose
periodic boundary conditions in the $y$ direction. The lattice extends
indefinitely to the left while it is terminated by a rough boundary
on the right. The boundary and bulk regions are bounded by green and
red dashed lines, respectively. }

\label{fig:RoughGrapheneBoundary}
\end{figure}

Although Eq.~(\ref{eq:GaussianRoughness}) describes the structure
of a continuous boundary, it is not immediately applicable to the
graphene lattice which is discrete and has a minimum length scale
associated with $a_{\text{cc}}$. Thus, it is necessary to introduce
a procedure that maps the continuous boundary described by $h(y)$
to the positions of the boundary C atoms. In our simulations, to construct
the atomistic realization of the boundary described by an instance
of $h(y),$we first divide the graphene lattice into hexagonal subunits
with each subunit centered at $(x_{\text{c}},y_{\text{c}})$ and consisting
of six C atoms. If the $x_{\text{c}}$ of a hexagonal subunit is positioned
to the left of the continuous boundary such that $x_{\text{c}}<h(y_{\text{c}})$,
then its six atoms are incorporated into the simulated boundary structure.
This procedure ensures that there are no dangling C-C bonds at the
boundary, i.e., each boundary atom is connected to at least two other
atoms. Figure~\ref{fig:RoughGrapheneBoundary} shows the realization
of a rough boundary of zigzag-edge graphene corresponding to an instance
of $h(y)$. After the positions of the C atoms are set, the entire
structure is optimized in GULP~\citep{JGale:MolSim03_gulp} using
the REBO potential~\citep{DWBrenner:JPCM02_REBO}. We obtain a value
of $a_{\text{cc}}=1.4203\text{Å}$ from the optimization of bulk graphene.
Although the REBO potential has not been optimized for the phonon
dispersion of graphene~\citep{LLindsay:PRB10_Optimized}, we use
it because it can accommodate a sufficiently large and stable rough
boundary needed for both the zigzag and armchair-edge boundary and
the primary objective of our work is to understand the effect of boundary
roughness on specularity at an atomistic level. 

In our simulations, the width (lineal cross-section) of the zigzag-edge
boundary and layers is $W=275.56\text{Å}$ in the $y$ direction.
This is the largest $W$ value that we can use in GULP to extract
the IFC values. We impose periodic boundary conditions for $h(y)$
in the $y$ direction so that $h(0)=h(W)$. For each combination of
$L$ and $\sigma$, we generate 20 instances of $h(y)$, which we
need for computing ensemble-averaged quantities $\langle\ldots\rangle$
in Eq.~(\ref{eq:ProbSpecReflection}), and we use each one to construct
an atomistic model of the boundary for the zigzag-edge graphene lattice.
For each atomistic boundary model, we calculate the interatomic force-constant
(IFC) matrices $\boldsymbol{H}_{\text{B}}$ and $\boldsymbol{H}_{\text{LB}}$
that describe the mass-normalized harmonic forces within the boundary
region (Fig.~\ref{fig:RoughGrapheneBoundary}) and the harmonic forces
between the bulk and the boundary region. 

\begin{figure*}
\includegraphics[scale=0.5]{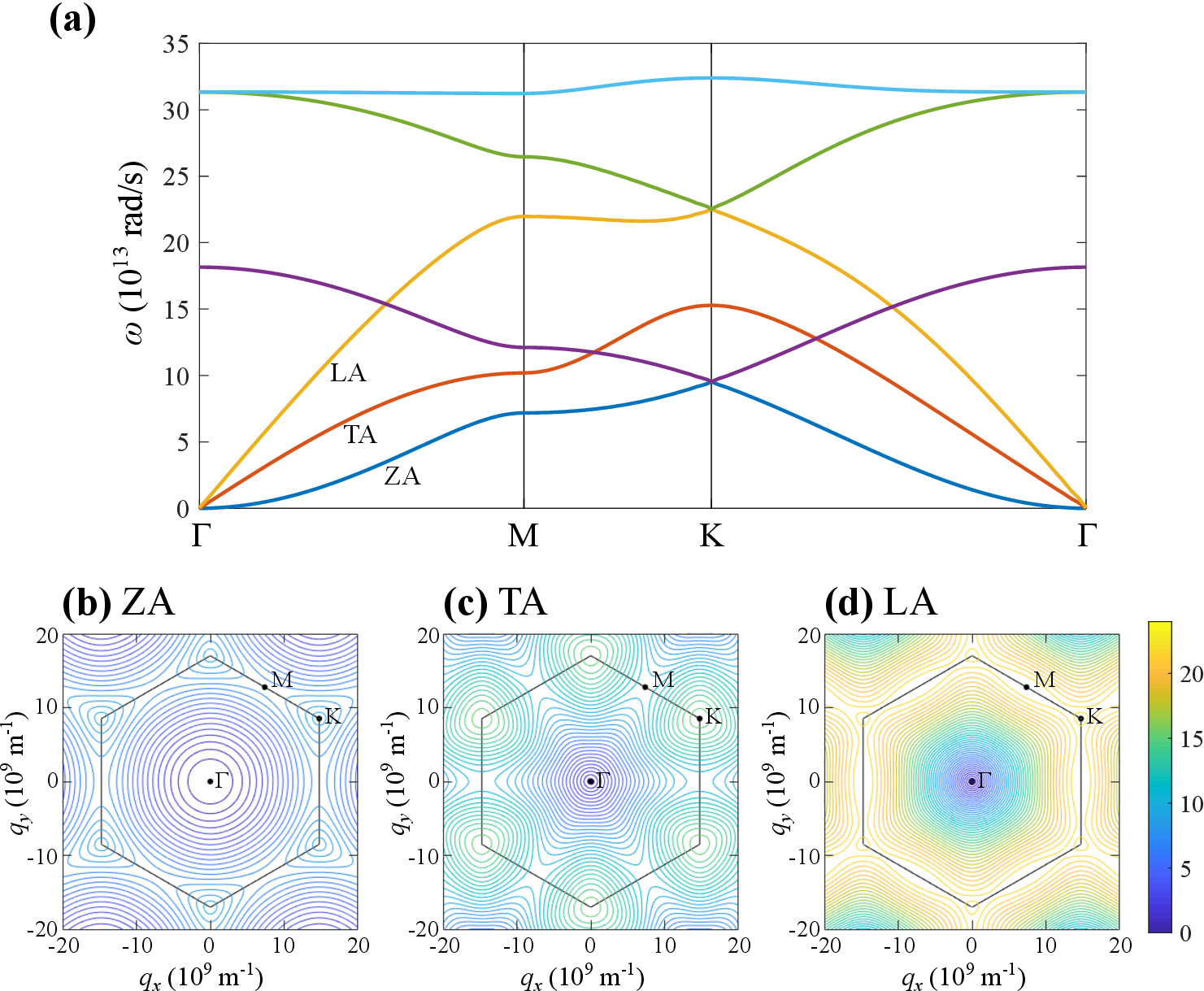}

\caption{(a) Plot of the bulk phonon dispersion between the symmetry points
($\Gamma$, $\text{K}$ and $\text{M}$) in the Brillouin Zone of
zigzag-edge graphene. The dispersion curves are calculated using the
IFC matrices obtained with the REBO potential. The different acoustic
phonon branches are identified using the method in Ref.~\citep{CGan:JPC21_Complementary}
and colored according to their polarization. In the long-wavelength
limit near the $\Gamma$ point, the LA and TA branches exhibit a linear
dispersion ($\omega\propto q$) while the ZA branch has a quadratic
dispersion ($\omega\propto q^{2}$). Panels (b)-(d) show the two-dimensional
contour plots of the dispersion in reciprocal space for the ZA, TA
and LA phonons in zigzag-edge graphene with $\omega$ indicated in
color. The Brillouin Zone boundary is indicated by the solid gray
lines. The frequency contours are drawn in intervals of $\Delta\omega=0.5\times10^{13}$
rad/s. }

\label{fig:BulkGraphenePhononDispersion}
\end{figure*}

Figure~\ref{fig:BulkGraphenePhononDispersion}(a) shows the phonon
dispersion curves for graphene, with each phonon branch distinctly
color-coded, using the interatomic force constants generated in GULP
after the structure for bulk graphene is optimized. The identification
and labeling of the phonon branch or polarization for each eigenmode
is carried out by using the method described in Ref.~\citep{CGan:JPC21_Complementary}.
In our work, we ignore the optical phonons and limit the scope in
our study of boundary roughness scattering to the acoustic phonons,
which have three distinct branches in graphene, because the Ogilvy
formula from Eq.~(\ref{eq:ProbOgilvyFormula}) is only applicable
to the acoustic waves. Figure~\ref{fig:BulkGraphenePhononDispersion}(b)
to (d) show the two-dimensional distribution of the eigenmode frequency
$\omega$ over the first Brillouin Zone as a function of the wave
vector $\bm{q}$, which we compute using the REBO potential, for the
flexural acoustic (ZA), transverse acoustic (TA), and longitudinal
acoustic (LA) phonons. In the long-wavelength limit near the $\Gamma$
point, the LA and TA branches exhibit a linear dispersion ($\omega\propto q$)
and have a well-defined wave speed while the ZA branch has a quadratic
dispersion ($\omega\propto q^{2}$). The slope for the LA phonon branch
is also significantly greater than the slope for the TA phonon branch
with the average longitudinal wave speed $c_{\text{L}}$ ($19.6$
km/s) nearly twice the average transverse wave speed $c_{\text{T}}$
($10.7$ km/s), the values of which we extract from Fig.~\ref{fig:BulkGraphenePhononDispersion}(a).

\subsection{Methodology for $S$ matrix calculation}

To describe the elastic scattering of phonons, we adopt the extended
AGF method~\citep{ZYOng:PRB15_Efficient,LYang:PRB18_Phonon} which
is developed in Refs.~\citep{ZYOng:JAP18_Tutorial,ZYOng:PRB18_Atomistic}
to describe mode-resolved transmission and reflection, and it has
been used to characterize diffuse phonon scattering by graphene grain
boundaries~\citep{ZYOng:PRB20_Structure,ZYOng:EPL21_Specular}. The
reader may skip this part of the paper and proceed directly to Sec.~\ref{sec:SimulationResults}
as the details of the calculation given in Sec.~\ref{subsec:ExtractInputSubmatrices}
to \ref{subsec:ComputeScatteringAmplitudes} are not necessary for
understanding the results discussed in Sec.~\ref{sec:SimulationResults}
although we give an overview of the AGF method here. The inputs for
the extended AGF method are the mass-normalized IFC submatrices ($\boldsymbol{H}_{0,0}$
and $\boldsymbol{H}_{0,1}$) associated with the bulk and the boundary
region ($\boldsymbol{H}_{\text{B}}$ and $\boldsymbol{H}_{\text{LB}}$)
as shown in Fig.~\ref{fig:RoughGrapheneBoundary}. For each boundary
structure, we use these four input submatrices to compute the frequency-dependent
$S$ matrix for each $\omega$ over the frequency range from $\omega=0.5\times10^{13}$
to $2.2\times10^{14}$ rad/s at intervals of $\Delta\omega=0.5\times10^{13}$
rad/s, with the upper bound of this range limited by the highest possible
LA phonon frequency at the $K$ point in Fig.~\ref{fig:BulkGraphenePhononDispersion}(a). 

At each $\omega$ step, all the possible incoming phonon modes are
computed in the AGF method, with the polarization and wave vector
of each mode labeled $\nu_{i}\boldsymbol{q}_{i}$. All of the possible
outgoing modes are also similarly computed, with each labeled $\nu_{r}\boldsymbol{q}_{r}$.
We then extract the target matrix elements $S_{\sigma}(\nu_{r}\boldsymbol{q}_{r},\nu_{i}\boldsymbol{q}_{i})_{q_{r}^{\parallel}=q_{i}^{\parallel}}$
for the ensemble of boundary structures corresponding to a $\sigma,L$
combination and compute the specularity parameter or probability of
specular reflection $p_{\sigma}(\nu_{r}\boldsymbol{q}_{r},\nu_{i}\boldsymbol{q}_{i})$
as given in Eq.~(\ref{eq:ProbSpecReflection}). The key formulas
for computing the $S$ matrix are given in Eqs.~(\ref{subeqs:SurfaceGreensFunctions})
to (\ref{eq:rLLmatrices}). 

\subsubsection{Extraction of input submatrices in bulk and boundary structure \label{subsec:ExtractInputSubmatrices}}

In the bulk graphene lattice, the atoms can be arranged as a periodic
array of layers in the direction perpendicular to the boundary so
that the overall IFC matrix can be expressed in the block-tridiagonal
form
\begin{equation}
\mathbf{H}_{\text{bulk}}=\left(\begin{array}{ccccc}
\ddots & \ddots\\
\ddots & \boldsymbol{H}_{-1,-1} & \boldsymbol{H}_{-1,0}\\
 & \boldsymbol{H}_{-1,0}^{\dagger} & \boldsymbol{H}_{0,0} & \boldsymbol{H}_{0,1}\\
 &  & \boldsymbol{H}_{0,1}^{\dagger} & \boldsymbol{H}_{1,1} & \ddots\\
 &  &  & \ddots & \ddots
\end{array}\right)\ ,\label{eq:TotalGrapheneIFC}
\end{equation}
where the submatrix $\boldsymbol{H}_{m,n}$ represents the mass-normalized
IFC coupling of layer $n$ to layer $m$ and $\boldsymbol{H}_{m,n}^{\dagger}=\boldsymbol{H}_{n,m}$.
The short-range interatomic forces imply that only neighboring layers
are coupled, i.e., only the submatrices $\boldsymbol{H}_{n,n-1}$,
$\boldsymbol{H}_{n,n}$ and $\boldsymbol{H}_{n,n+1}$ have non-zero
matrix elements. The translational symmetry means that each layer
is identical so that $\boldsymbol{H}_{0,0}=\boldsymbol{H}_{1,1}=\ldots$
and $\boldsymbol{H}_{-1,0}=\boldsymbol{H}_{0,1}=\ldots$. Hence, only
two unique submatrices $\boldsymbol{H}_{0,0}$ and $\boldsymbol{H}_{0,1}$
are needed to construct $\mathbf{H}_{\text{bulk}}$. The IFC matrix
for the semi-infinite graphene system with a rough boundary has the
form 
\begin{equation}
\mathbf{H}_{\text{boundary}}=\left(\begin{array}{ccccc}
\ddots & \ddots\\
\ddots & \boldsymbol{H}_{0,0} & \boldsymbol{H}_{0,1}\\
 & \boldsymbol{H}_{0,1}^{\dagger} & \boldsymbol{H}_{0,0} & \boldsymbol{H}_{0,1}\\
 &  & \boldsymbol{H}_{0,1}^{\dagger} & \boldsymbol{H}_{0,0} & \boldsymbol{H}_{\text{LB}}\\
 &  &  & \boldsymbol{H}_{\text{LB}}^{\dagger} & \boldsymbol{H}_{\text{B}}
\end{array}\right)\label{eq:GrapheneBoundaryIFC}
\end{equation}
where $\boldsymbol{H}_{\text{B}}$ describes the IFC coupling within
the boundary region shown in Fig.~\ref{fig:RoughGrapheneBoundary}
while $\boldsymbol{H}_{\text{LB}}$ describes the IFC coupling between
the bulk and the boundary region. The correspondence of the individual
IFC submatrices ($\boldsymbol{H}_{0,0}$, $\boldsymbol{H}_{0,1}$,
$\boldsymbol{H}_{\text{LB}}$ and $\boldsymbol{H}_{\text{B}}$) to
the arrangement of the layers in the simulated system is shown in
Fig.~\ref{fig:RoughGrapheneBoundary}. We extract the first pair
of submatrices $\boldsymbol{H}_{0,0}$ and $\boldsymbol{H}_{0,1}$
from $\mathbf{H}_{\text{bulk}}$ and the second pair $\boldsymbol{H}_{\text{LB}}$
and $\boldsymbol{H}_{\text{B}}$ from $\mathbf{H}_{\text{boundary}}$. 

\subsubsection{Computation of Bloch matrices and eigenmodes associated with bulk
graphene \label{subsec:ComputeBlochMatrices}}

To study the elastic scattering of phonons, we limit the dynamics
to a fixed frequency $\omega$. At each frequency $\omega$, to find
the eigenmodes associated with the translational symmetry of the bulk
region in the longitudinal $x$ direction, we need to use the submatrices
$\boldsymbol{H}_{0,0}$ and $\boldsymbol{H}_{0,1}$ from Sec.~\ref{subsec:ExtractInputSubmatrices}.
We first define the frequency-dependent surface Green's function matrices
$\bm{g}_{\text{L},-}^{\text{ret}}(\omega)$ and $\boldsymbol{g}_{\text{L},-}^{\text{adv}}(\omega)$:
\begin{subequations}

\begin{equation}
\bm{g}_{\text{L},-}^{\text{ret}}(\omega)=[(\omega+i0^{+})^{2}\bm{I}-\boldsymbol{H}_{0,0}-\boldsymbol{H}_{0,1}^{\dagger}\bm{g}_{\text{L},-}^{\text{ret}}\boldsymbol{H}_{0,1}]^{-1}\ ,
\end{equation}
\begin{equation}
\boldsymbol{g}_{\text{L},-}^{\text{adv}}(\omega)=\boldsymbol{g}_{\text{L},-}^{\text{ret}}(\omega)^{\dagger}\ .
\end{equation}
\label{subeqs:SurfaceGreensFunctions}\end{subequations}Given Eq.~(\ref{subeqs:SurfaceGreensFunctions}),
we define the \emph{Bloch matrices} $\boldsymbol{F}_{\text{L},-}^{\text{ret}}(\omega)$
and $\boldsymbol{F}_{\text{L},-}^{\text{adv}}(\omega)$: \begin{subequations}
\begin{equation}
[\boldsymbol{F}_{\text{L},-}^{\text{ret}}(\omega)]{}^{-1}=\boldsymbol{g}_{\text{L},-}^{\text{ret}}(\omega)\boldsymbol{H}_{0,1}\ ,\label{eq:LeftBlochMatricesRet}
\end{equation}
\begin{equation}
[\boldsymbol{F}_{\text{L},-}^{\text{adv}}(\omega)]{}^{-1}=\boldsymbol{g}_{\text{L},-}^{\text{adv}}(\omega)\boldsymbol{H}_{0,1}\ \label{eq:LeftBlochMatricesAdv}
\end{equation}
\label{subeqs:BlochMatrices}\end{subequations}which describe the
translational symmetry of the eigenmodes. This allows us to determine
the Bloch eigenmode matrices $\boldsymbol{U}_{\text{L},-}^{\text{ret}}(\omega)$
and $\boldsymbol{U}_{\text{L},-}^{\text{adv}}(\omega)$: \begin{subequations}
\begin{align}
[\boldsymbol{F}_{\text{L},-}^{\text{ret}}(\omega)]^{-1}\boldsymbol{U}_{\text{L},-}^{\text{ret}}(\omega) & =\boldsymbol{U}_{\text{L},-}^{\text{ret}}(\omega)[\boldsymbol{\Lambda}_{\text{L},-}^{\text{ret}}(\omega)]^{-1}\ ,\label{eq:LeftBlochEigenmodesRet}\\{}
[\boldsymbol{F}_{\text{L},-}^{\text{adv}}(\omega)]^{-1}\boldsymbol{U}_{\text{L},-}^{\text{adv}}(\omega) & =\boldsymbol{U}_{\text{L},-}^{\text{adv}}(\omega)[\boldsymbol{\Lambda}_{\text{L},-}^{\text{adv}}(\omega)]^{-1}\ \label{eq:LeftBlochEigenmodesAdv}
\end{align}
\label{subeqs:BlochEigenmodes}\end{subequations}associated with
the outgoing leftward-propagating ($\boldsymbol{U}_{\text{L},-}^{\text{ret}}$)
and incoming rightward-propagating ($\boldsymbol{U}_{\text{L},-}^{\text{adv}}$)
phonon eigenmodes at frequency $\omega$. The eigenvalue matrices
$\boldsymbol{\Lambda}_{\text{L},-}^{\text{ret}}(\omega)$ and $\boldsymbol{\Lambda}_{\text{L},-}^{\text{adv}}(\omega)$
have the diagonal form
\[
\boldsymbol{\Lambda}_{\text{L},-}^{\text{ret}}(\omega)=\left(\begin{array}{ccc}
e^{ik_{1}^{\perp}a}\\
 & e^{ik_{2}^{\perp}a}\\
 &  & \ddots
\end{array}\right)
\]
where $k_{1}^{\perp}$, $k_{2}^{\perp}$, $\ldots$ are the `folded'
wave vector components in the longitudinal $x$ direction for the
phonon eigenmodes at frequency $\omega$. Because of the periodic
boundary condition in the transverse ($y$) direction, we can also
associate each phonon eigenmode with a `folded' wave vector component
in the transverse direction and determine the corresponding set of
transverse wave vectors $k_{1}^{\parallel}$, $k_{2}^{\parallel}$,
$\dots$. Using the zone-unfolding method from Ref.~\citep{ZYOng:PRB18_Atomistic},
we can map each of the `folded' 2D wave vectors, e.g. $\bm{k}=(k^{\perp},k^{\parallel})$,
to an unfolded 2D wave vector $\bm{q}=(q_{x},q_{y})$ corresponding
to a unique wave vector within the first Brillouin Zone. Therefore,
we can associate each phonon eigenmode with an unfolded wave vector
and thus the phonon eigenmode matrix from Eq.~(\ref{subeqs:BlochEigenmodes})
can be expressed as \begin{subequations}
\begin{equation}
\boldsymbol{U}_{\text{L},-}^{\text{ret}}(\omega)=[\boldsymbol{u}_{\text{L},-}^{\text{ret}}(\bm{q}_{1}),\boldsymbol{u}_{\text{L},-}^{\text{ret}}(\bm{q}_{2}),\ldots]\label{eq:UnfoldedLeftBlochEigenmodesRet}
\end{equation}
where $\boldsymbol{u}_{\text{L},-}^{\text{ret}}(\bm{q})$ is the column
vector corresponding to the outgoing phonon eigenmode with wave vector
$\bm{q}$, which is real for bulk phonon modes and complex for evanescent
modes, at frequency $\omega$. In addition, a unique phonon polarization
$\nu$ can be associated with the phonon eigenmode for $\bm{q}$ and
$\omega$, using the method described in Ref.~\citep{CGan:JPC21_Complementary}.
Similarly, we can write 
\begin{equation}
\boldsymbol{U}_{\text{L},-}^{\text{adv}}(\omega)=[\boldsymbol{u}_{\text{L},-}^{\text{adv}}(\bm{q}_{1}^{\prime}),\boldsymbol{u}_{\text{L},-}^{\text{adv}}(\bm{q}_{2}^{\prime}),\ldots]\label{eq:UnfoldedLeftBlochEigenmodesAdv}
\end{equation}
\label{subeqs:UnfoldedBlochEigenmodes}\end{subequations}where $\boldsymbol{u}_{\text{L},-}^{\text{adv}}(\bm{q}^{\prime})$
is the column vector corresponding to the incoming phonon eigenmode
with wave vector $\bm{q}^{\prime}$.

Given the Bloch eigenmodes from Eq.~(\ref{subeqs:UnfoldedBlochEigenmodes}),
we also define their associated eigenvelocity matrices $\boldsymbol{V}_{\text{L},-}^{\text{ret}}(\omega)$
and $\boldsymbol{V}_{\text{L},-}^{\text{adv}}(\omega)$: \begin{subequations}
\begin{align}
\boldsymbol{V}_{\text{L},-}^{\text{ret}}(\omega) & =-\frac{ia_{\text{L}}}{2\omega}(\boldsymbol{U}_{\text{L},-}^{\text{ret}})^{\dagger}\boldsymbol{H}_{0,1}^{\dagger}[\boldsymbol{g}_{\text{L},-}^{\text{ret}}-(\boldsymbol{g}_{\text{L},-}^{\text{ret}})^{\dagger}]\boldsymbol{H}_{0,1}\boldsymbol{U}_{\text{L},-}^{\text{ret}}\label{eq:LeftBlochEigenvelRet}\\
\boldsymbol{V}_{\text{L},-}^{\text{adv}}(\omega) & =-\frac{ia_{\text{L}}}{2\omega}(\boldsymbol{U}_{\text{L},-}^{\text{adv}})^{\dagger}\boldsymbol{H}_{0,1}^{\dagger}[\boldsymbol{g}_{\text{L},-}^{\text{adv}}-(\boldsymbol{g}_{\text{L},-}^{\text{adv}})^{\dagger}]\boldsymbol{H}_{0,1}\boldsymbol{U}_{\text{L},-}^{\text{adv}}\label{eq:LeftBlochEigenvelAdv}
\end{align}
\label{subeqs:BlochEigenvelocities}\end{subequations}where $a_{\text{L}}$
is the interlayer distance. The eigenvelocity matrices from Eq.~(\ref{subeqs:BlochEigenvelocities})
have the diagonal form
\begin{align*}
\boldsymbol{V}_{\text{L},-}^{\text{ret}}(\omega) & =\left(\begin{array}{ccc}
v_{\text{L},-}^{\text{ret}}(\bm{q}_{1})\\
 & v_{\text{L},-}^{\text{ret}}(\bm{q}_{2})\\
 &  & \ddots
\end{array}\right)\\
\boldsymbol{V}_{\text{L},-}^{\text{adv}}(\omega) & =\left(\begin{array}{ccc}
v_{\text{L},-}^{\text{adv}}(\bm{q}_{1}^{\prime})\\
 & v_{\text{L},-}^{\text{adv}}(\bm{q}_{2}^{\prime})\\
 &  & \ddots
\end{array}\right)
\end{align*}
where $v_{\text{L},-}^{\text{ret}}(\bm{q})$ and $v_{\text{L},-}^{\text{adv}}(\bm{q})$
are the longitudinal velocity components for the outgoing and incoming
phonon modes with wave vector $\bm{q}$, respectively. Because the
incoming phonon modes are traveling rightward towards the boundary
while the outgoing phonon modes are traveling leftward away from the
boundary, we have $v_{\text{L},-}^{\text{ret}}\leq0$ and $v_{\text{L},-}^{\text{adv}}\geq0$.

\subsubsection{Computation of reflection matrix and scattering amplitudes \label{subsec:ComputeScatteringAmplitudes}}

Finally, to compute the scattering amplitudes of the reflected phonons,
we need the Green's function submatrix for the boundary region in
the semi-infinite SLG sheet,
\begin{equation}
\bm{G}_{\text{B}}^{\text{ret}}(\omega)=[(\omega+i0^{+})^{2}\bm{I}_{\text{B}}-\boldsymbol{H}_{\text{B}}-\bm{H}_{\text{LB}}^{\dagger}\bm{g}_{\text{L},-}^{\text{ret}}\bm{H}_{\text{LB}}]^{-1}\label{eq:RetGreensFuncBoundaryRegion}
\end{equation}
which requires all four input submatrices ($\boldsymbol{H}_{0,0}$,
$\boldsymbol{H}_{0,1}$, $\boldsymbol{H}_{\text{LB}}$ and $\boldsymbol{H}_{\text{B}}$).
We also define 
\begin{align}
\bm{Q}_{\text{L}}(\omega)= & (\omega+i0^{+})^{2}\bm{I}-\boldsymbol{H}_{0,0}-\boldsymbol{H}_{0,1}^{\dagger}\bm{g}_{\text{L},-}^{\text{ret}}\boldsymbol{H}_{0,1}\label{eq:RetBulkGreensFunc}\\
 & -\boldsymbol{H}_{0,1}\bm{g}_{\text{L},+}^{\text{ret}}\boldsymbol{H}_{0,1}^{\dagger}\nonumber 
\end{align}
where $\bm{g}_{\text{L},+}^{\text{ret}}(\omega)=[(\omega+i0^{+})^{2}\bm{I}-\boldsymbol{H}_{0,0}-\boldsymbol{H}_{0,1}\bm{g}_{\text{L},+}^{\text{ret}}(\omega)\boldsymbol{H}_{0,1}^{\dagger}]^{-1}$. 

Given Eqs.~(\ref{eq:RetGreensFuncBoundaryRegion}) and (\ref{eq:RetBulkGreensFunc}),
the reflection matrix, which relates the amplitude of the outgoing
phonon flux to the incoming phonon flux, is given by
\begin{align}
\boldsymbol{r}_{\text{LL}}(\omega)= & \frac{2i\omega}{a_{\text{L}}}(\boldsymbol{V}_{\text{L},-}^{\text{ret}})^{1/2}(\boldsymbol{U}_{\text{L},-}^{\text{ret}})^{-1}(\boldsymbol{G}_{\text{L}}^{\text{ret}}-\boldsymbol{Q}_{\text{L}}^{-1})\nonumber \\
 & (\boldsymbol{U}_{\text{L},-}^{\text{adv}}{}^{\dagger})^{-1}(\boldsymbol{V}_{\text{L},-}^{\text{adv}})^{1/2}\label{eq:ReflectionMatrix}
\end{align}
where $\boldsymbol{G}_{\text{L}}^{\text{ret}}(\omega)=\boldsymbol{g}_{\text{L},-}^{\text{ret}}+\boldsymbol{g}_{\text{L},-}^{\text{ret}}\boldsymbol{H}_{\text{LB}}\boldsymbol{G}_{\text{B}}^{\text{ret}}\boldsymbol{H}_{\text{LB}}^{\dagger}\boldsymbol{g}_{\text{L},-}^{\text{ret}}$.
We obtain the reduced reflection matrix $\overline{\boldsymbol{r}}_{\text{LL}}$
from Eq.~(\ref{eq:ReflectionMatrix}) by eliminating the matrix columns
and rows associated with the evanescent modes. The matrix $\overline{\boldsymbol{r}}_{\text{LL}}$
is an $N\times N$ matrix of the form: 
\begin{equation}
\overline{\boldsymbol{r}}_{\text{LL}}=\left(\begin{array}{ccc}
S(\nu_{1}\bm{q}_{1},\nu_{1}^{\prime}\bm{q}_{1}^{\prime}) & \ldots & S(\nu_{1}\bm{q}_{1},\nu_{N}^{\prime}\bm{q}_{N}^{\prime})\\
\vdots & \ddots & \vdots\\
S(\nu_{N}\bm{q}_{N},\nu_{1}^{\prime}\bm{q}_{1}^{\prime}) & \ldots & S(\nu_{N}\bm{q}_{N},\nu_{N}^{\prime}\bm{q}_{N}^{\prime})
\end{array}\right)\label{eq:rLLmatrices}
\end{equation}
where $S(\nu_{n}\boldsymbol{q}_{n},\nu_{m}^{\prime}\boldsymbol{q}_{m}^{\prime})$
is the complex scattering amplitude between the incoming $\nu_{m}^{\prime}\boldsymbol{q}_{m}^{\prime}$
phonon mode and the outgoing $\nu_{n}\boldsymbol{q}_{n}$ phonon mode
from Eq.~\ref{subeqs:UnfoldedBlochEigenmodes}. We note here that
because the width of the system $W$ is finite, the transverse components
of the wave vectors for the incoming and outgoing modes, $q_{i}^{\parallel}$
and $q_{r}^{\parallel}$, are discretized such that $q^{\parallel}=\frac{2\pi n}{W}$
where $n$ in an integer. To compute the specularity parameter $p_{\sigma}(\nu_{r}\boldsymbol{q}_{r},\nu_{i}\boldsymbol{q}_{i})$
from Eq.~(\ref{eq:ProbSpecReflection}), we take $\nu_{r}$, $\nu_{i}$
and $\bm{q}_{i}$ as inputs and find the matching matrix element $S(\nu_{r}\boldsymbol{q}_{r},\nu_{i}\boldsymbol{q}_{i})$
from Eq.~(\ref{eq:rLLmatrices}) for which $q_{r}^{\parallel}=q_{i}^{\parallel}$
to obtain $S_{\sigma}(\nu_{r}\boldsymbol{q}_{r},\nu_{i}\boldsymbol{q}_{i})_{q_{r}^{\parallel}=q_{i}^{\parallel}}$
for that particular boundary.

\section{Simulation results and discussion \label{sec:SimulationResults}}

\subsection{Attenuation parameter $\chi$}

We can quantify the attenuation of the specular reflection by using
the ratio
\begin{equation}
\zeta_{\sigma}(\nu_{r}\boldsymbol{q}_{r},\nu_{i}\boldsymbol{q}_{i})=p_{\sigma}(\nu_{r}\boldsymbol{q}_{r},\nu_{i}\boldsymbol{q}_{i})/p_{0}(\nu_{r}\boldsymbol{q}_{r},\nu_{i}\boldsymbol{q}_{i})\label{eq:SpecularityAttenuationRatio}
\end{equation}
where $p_{0}(\nu_{r}\boldsymbol{q}_{r},\nu_{i}\boldsymbol{q}_{i})\geq\epsilon$
and $\epsilon$ is equal to a small numerical constant that corresponds
to the minimum nonzero probability of specular reflection ($q_{r}^{\parallel}=q_{i}^{\parallel}$)
when $\sigma=0$. This is done to exclude scattering processes in
which specular reflection is forbidden (e.g. $\text{TA}\rightarrow\text{LA}$
scattering for $\theta_{i}>\theta_{\text{c}}$). In our case, we obtain
reasonable results for $\epsilon=10^{-6}$. To compute $p_{\sigma}(\nu_{r}\boldsymbol{q}_{r},\nu_{i}\boldsymbol{q}_{i})$
for each combination of $L$ and $\sigma$, we use an ensemble of
$N=20$ realizations of the graphene rough boundary for the ensemble
averages in Eq.~(\ref{eq:ProbSpecReflection}).

If the Ogilvy formula from Eq.~(\ref{eq:ProbOgilvyFormula}) holds,
it implies that $\zeta_{\sigma}(\nu_{r}\boldsymbol{q}_{r},\nu_{i}\boldsymbol{q}_{i})=\exp[-\sigma^{2}(|q_{i}^{\perp}|+|q_{r}^{\perp}|)^{2}]$
or $|\log[\zeta_{\sigma}(\nu_{r}\boldsymbol{q}_{r},\nu_{i}\boldsymbol{q}_{i})]|^{1/2}\propto\sigma$.
Therefore, we define the dimensionless \emph{attenuation parameter
}as $\chi(\nu_{r}\boldsymbol{q}_{r},\nu_{i}\boldsymbol{q}_{i})=|\log\zeta_{\sigma}(\nu_{r}\boldsymbol{q}_{r},\nu_{i}\boldsymbol{q}_{i})|^{1/2}$
or\emph{ }
\begin{equation}
\chi(\nu_{r}\boldsymbol{q}_{r},\nu_{i}\boldsymbol{q}_{i})=\left|\log\left[\frac{p_{\sigma}(\nu_{r}\boldsymbol{q}_{r},\nu_{i}\boldsymbol{q}_{i})}{p_{0}(\nu_{r}\boldsymbol{q}_{r},\nu_{i}\boldsymbol{q}_{i})}\right]\right|^{1/2}\ \label{eq:AttenuationParameter}
\end{equation}
to characterize the degree of attenuation from boundary roughness
scattering in our simulation results. If the Ogilvy formula holds,
then we have $\zeta_{\sigma}(\nu_{r}\boldsymbol{q}_{r},\nu_{i}\boldsymbol{q}_{i})=\exp[-\chi(\nu_{r}\boldsymbol{q}_{r},\nu_{i}\boldsymbol{q}_{i})^{2}]$
and we may regards Eq.~(\ref{eq:AttenuationParameter}) as a generalization
of the Rayleigh roughness parameter $\Sigma$ from Eq.~(\ref{eq:RayleighFormula})
since $\Sigma=\frac{1}{2}\chi$. If there is no attenuation, then
$\chi=0$; if specular reflection is totally eliminated, then $\chi=\infty$.
For Eq.~(\ref{eq:AttenuationParameter}), we have $|q_{r}^{\perp}|=|q_{i}^{\perp}|$
in the absence of mode conversion ($\nu_{r}=\nu_{i}$) and $|q_{r}^{\perp}|\neq|q_{i}^{\perp}|$
otherwise ($\nu_{r}\neq\nu_{i}$). For convenience, we define the
average of the normal component of the incident and reflected wave
vectors as
\begin{equation}
Q_{x}=\frac{|q_{i}^{\perp}|+|q_{r}^{\perp}|}{2}\label{eq:DefinitionQxVariable}
\end{equation}
to obtain the expression $\zeta_{\sigma}(\nu_{r}\boldsymbol{q}_{r},\nu_{i}\boldsymbol{q}_{i})=\exp[-4\sigma^{2}Q_{x}^{2}]$
or $\chi(\nu_{r}\boldsymbol{q}_{r},\nu_{i}\boldsymbol{q}_{i})=2\sigma Q_{x}$
from Eq.~(\ref{eq:ProbOgilvyFormula}). We note that $Q_{x}$ is
a function of $\bm{q}_{i}$ and hence $q_{x}$ since $q_{i}^{\perp}=q_{x}$
and $q_{r}^{\perp}$ is uniquely determined by $q_{i}^{\perp}$ when
$\nu_{r}$ and $\nu_{i}$ are given. If there is no mode conversion
($\nu_{r}=\nu_{i}$), then $Q_{x}=|q_{x}|$ because $q_{r}^{\perp}=q_{i}^{\perp}$.
Conversely, if there is mode conversion ($\nu_{r}\neq\nu_{i}$), then
$Q_{x}\neq|q_{x}|$. 

\subsection{Specular reflection for no boundary roughness ($\sigma=0$)}

To apply Eq.~(\ref{eq:ProbOgilvyFormula}), we first compute the
probability of specular reflectance $p_{0}(\nu_{r}\boldsymbol{q}_{r},\nu_{i}\boldsymbol{q}_{i})$,
which will serve as the baseline case when there is no boundary roughness
($\sigma=0$ or $\mathcal{T}=0$), with and without mode conversion.
The $p_{0}(\nu_{r}\boldsymbol{q}_{r},\nu_{i}\boldsymbol{q}_{i})$
data also tells us which specular scattering processes with or without
mode conversion are allowed and which ones are forbidden.

\begin{figure*}
\includegraphics[scale=0.4]{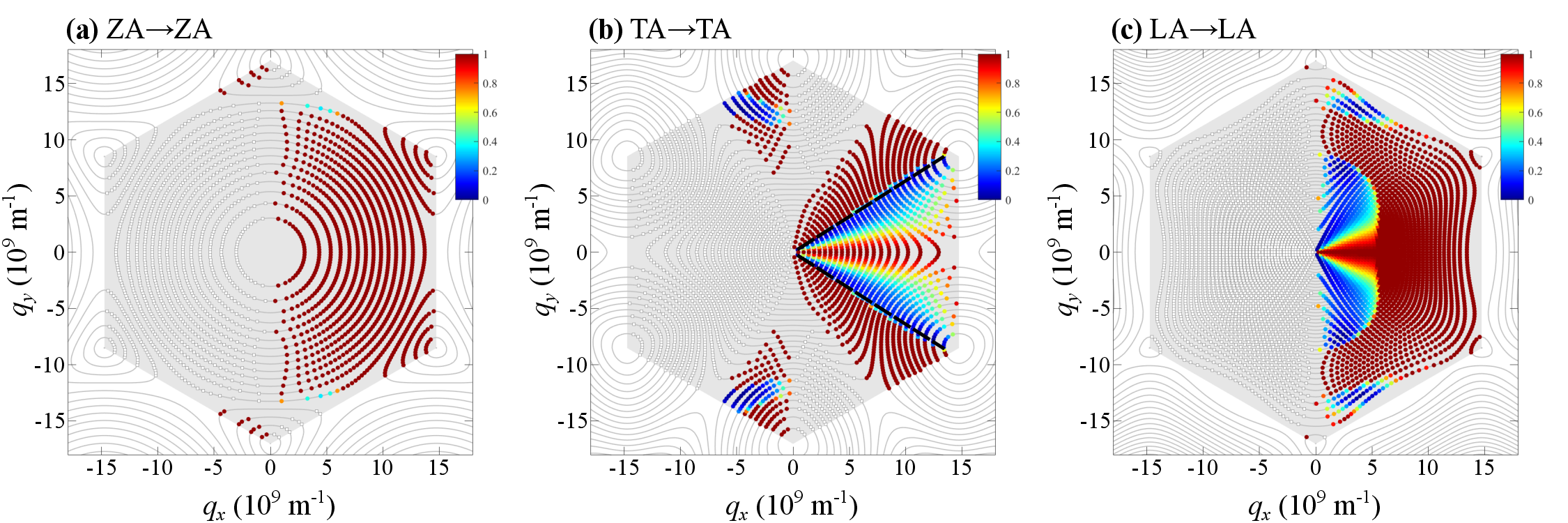}

\caption{Plot of the probability of specular reflectance $p_{0}(\nu_{r}\boldsymbol{q}_{r},\nu_{i}\boldsymbol{q}_{i})$
with no mode conversion ($\nu_{r}=\nu_{i}$ and $|q_{r}^{\perp}|=|q_{i}^{\perp}|$)
by a perfect ($\sigma=0$) zigzag-edge graphene boundary for the (a)
ZA, (b) TA and (c) LA phonons distributed over the first Brillouin
Zone (gray-shaded region). Each solid circle at $\bm{q}_{i}=(q_{x},q_{y})$
in the Brillouin Zone corresponds to an incident rightward-propagating
phonon mode, with its color representing the value of $p_{0}(\nu_{r}\boldsymbol{q}_{r},\nu_{i}\boldsymbol{q}_{i})$.
Because only rightward-propagating phonon modes are indicated, only
half of the BZ is covered by the solid circles. The other half of
the BZ is covered by hollow circles corresponding to the leftward-propagating
phonon modes that cannot scatter with the boundary. The phonon modes
are located on the frequency contours (gray solid lines) drawn at
intervals of $\Delta\omega=0.5\times10^{13}$ rad/s like in Fig.~\ref{fig:BulkGraphenePhononDispersion}.
The critical angle $\theta_{\text{c}}$ for the mode conversion of
the TA phonons is indicated by the black dashed line in (b).}

\label{fig:BaseReflectanceNoMode}
\end{figure*}

In Fig.~\ref{fig:BaseReflectanceNoMode}, we plot $p_{0}(\nu_{r}\boldsymbol{q}_{r},\nu_{i}\boldsymbol{q}_{i})$
with no mode conversion, where $\nu_{r}=\nu_{i}$ and $|q_{r}^{\perp}|=|q_{i}^{\perp}|$,
for incident (a) ZA, (b) TA and (c) LA phonons in zigzag-edge graphene.
The data in Fig.~\ref{fig:BaseReflectanceNoMode}(a) to (c) correspond
to the $\text{ZA}\rightarrow\text{ZA}$, $\text{TA}\rightarrow\text{TA}$
and $\text{LA}\rightarrow\text{LA}$ scattering processes, respectively.
In Fig.~\ref{fig:BaseReflectanceNoMode}, each solid circle is located
at $(q_{x},q_{y})$ and is associated with an \emph{incoming} phonon
mode that has the wave vector $\bm{q}_{i}=(q_{x},q_{y})$ within the
first Brillouin Zone and a velocity component ($v_{x}$) that is directed
rightwards towards the boundary, i.e., $v_{x}>0$. The phonon modes
that have a leftward velocity component ($v_{x}<0$) are not shown
because they are associated with the \emph{outgoing} phonon modes
reflected by the boundary. Because the calculations of the $S$ matrices
are performed in steps of $\omega$, we obtain a set of $p_{0}(\nu_{r}\boldsymbol{q}_{r},\nu_{i}\boldsymbol{q}_{i})$
values for each $\omega$ value which we can see in Fig.~\ref{fig:BaseReflectanceNoMode}
where the data point for each phonon mode is located on one of the
frequency contour lines. 

For the ZA phonons {[}Fig.~\ref{fig:BaseReflectanceNoMode}(a){]},
the value of $p_{0}$ is almost uniformly equal to unity within the
BZ, indicating an absence of mode conversion in boundary scattering,
which we attribute to the planar symmetry of SLG. On the other hand,
for the TA phonons {[}Fig.~\ref{fig:BaseReflectanceNoMode}(b){]},
the value of $p_{0}$ deviates significantly from unity, indicating
the presence of $\text{TA}\rightarrow\text{LA}$ mode conversion,
when the angle of incidence $\theta_{i}$ is smaller than the critical
angle $\theta_{\text{c}}$, which is indicated by the black solid
line in Fig.~\ref{fig:BaseReflectanceNoMode}(b) and given by $\sin\theta_{\text{c}}=c_{\text{T}}/c_{\text{L}}$.
As $\theta_{i}$ approaches zero (normal incidence), the degree of
mode conversion for the TA phonons decreases and converges to zero
at $\theta_{i}=0$. When $\theta_{i}>\theta_{\text{c}}$, the value
of $p_{0}$ is unity almost everywhere in the BZ, indicating the absence
of mode conversion. For the LA phonons {[}Fig.~\ref{fig:BaseReflectanceNoMode}(c){]},
$\text{LA}\rightarrow\text{TA}$ mode conversion is not angle-limited
because $c_{\text{L}}>c_{\text{T}}$ but has a sharp frequency cutoff
because the process is limited by the maximum frequency of the TA
phonons at the edge of the BZ ($\omega=1.5\times10^{14}$ rad/s).
Like the case for TA phonons, the degree of mode conversion converges
to zero as $\theta_{i}$ approaches zero. 

\begin{figure*}
\includegraphics[scale=0.4]{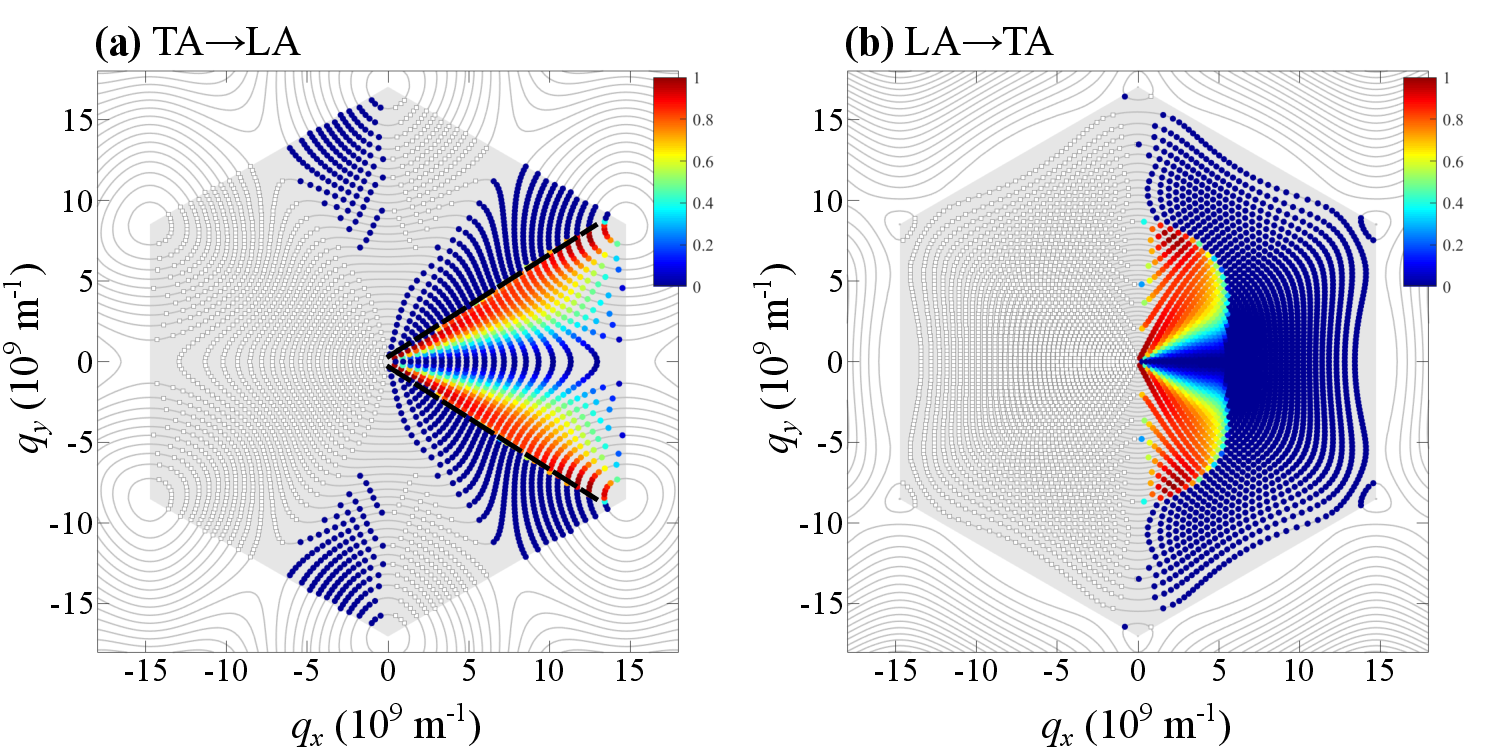}

\caption{Plot of the probability of specular reflectance $p_{0}(\nu_{r}\boldsymbol{q}_{r},\nu_{i}\boldsymbol{q}_{i})$
with only mode conversion ($\nu_{r}\protect\neq\nu_{i}$ and $|q_{r}^{\perp}|\protect\neq|q_{i}^{\perp}|$)
by a perfect ($\sigma=0$) zigzag-edge graphene boundary for the (a)
TA and (b) LA phonons distributed over the first Brillouin Zone (gray-shaded
region). Each solid circle at $\bm{q}_{i}=(q_{x},q_{y})$ in the Brillouin
Zone corresponds to an incident rightward-propagating phonon mode,
with its color representing the value of $p_{0}(\nu_{r}\boldsymbol{q}_{r},\nu_{i}\boldsymbol{q}_{i})$.
Like in Figs.~\ref{fig:BulkGraphenePhononDispersion} and \ref{fig:BaseReflectanceNoMode},
the frequency contours are drawn at intervals of $\Delta\omega=0.5\times10^{13}$
rad/s. The critical angle $\theta_{\text{c}}$ for the mode conversion
of the TA phonons is indicated by the black dashed line in (a).}

\label{fig:BaseReflectanceModeConv}
\end{figure*}

Similarly, in Fig.~\ref{fig:BaseReflectanceModeConv}, we plot $p_{0}(\nu_{r}\boldsymbol{q}_{r},\nu_{i}\boldsymbol{q}_{i})$
with only mode conversion ($\nu_{r}\neq\nu_{i}$ and $|q_{r}^{\perp}|\neq|q_{i}^{\perp}|$)
for the TA and LA phonons, with the data corresponding to the $\text{TA\ensuremath{\rightarrow}LA}$
and $\text{LA\ensuremath{\rightarrow}TA}$ scattering processes. We
do not plot $p_{0}(\nu_{r}\boldsymbol{q}_{r},\nu_{i}\boldsymbol{q}_{i})$
for the ZA phonons because they cannot undergo mode conversion, i.e.,
there are no $\text{ZA}\rightarrow\text{LA}$ or $\text{ZA}\rightarrow\text{TA}$
scattering processes, due to the planar symmetry of the SLG boundary.
As expected, the results in Fig.~\ref{fig:BaseReflectanceModeConv}(a)
and (b) are complementary and opposite to those in Fig.~\ref{fig:BaseReflectanceNoMode}(b)
and (c). The values of $p_{0}$ with only mode conversion are non-zero
for the TA phonons when $\theta_{i}<\theta_{\text{c}}$ and converge
to zero as $\theta_{i}$ approaches zero. Similarly, the values of
$p_{0}$ with only mode conversion are non-zero for the LA phonons
when $\theta_{i}<\pi/2$ and converge to zero as $\theta_{i}$ approaches
zero.

\subsection{Specular reflection for finite boundary roughness ($\sigma\protect\neq0$)}

To investigate the effects of boundary roughness scattering, we compute
$p_{\sigma}(\nu_{r}\boldsymbol{q}_{r},\nu_{i}\boldsymbol{q}_{i})$
with and without mode conversion for the ZA, TA and LA phonons and
boundary structures at different values of the lateral correlation
length $L$ and boundary roughness $\sigma$. We quantify the effects
by analyzing the reciprocal-space distribution of $p_{\sigma}(\nu_{r}\boldsymbol{q}_{r},\nu_{i}\boldsymbol{q}_{i})$
and $\chi(\nu_{r}\boldsymbol{q}_{r},\nu_{i}\boldsymbol{q}_{i})$ from
Eqs.~(\ref{eq:ProbSpecReflection}) and (\ref{eq:AttenuationParameter}),
respectively. Three sets of boundary structures with different $\sigma$
and $L$ values are used in our simulations. The first set has $\sigma=0.5R_{0}$
and $L=L_{0}$ and describes boundary structures with a small boundary
roughness and small correlation length. The second set has $\sigma=1.5R_{0}$
and $L=L_{0}$ and describes boundary structures with a large boundary
roughness and small correlation length. We use the contrast between
the first and second set to investigate the change in $\chi$ when
the boundary roughness is increased. The third set $\sigma=0.5R_{0}$
and $L=8L_{0}$ and describes boundary structures with a small boundary
roughness and large correlation length. The third set describes a
much smoother boundary and is used to investigate the change in $\chi$
when the lateral correlation length is larger. 

\subsubsection{Small lateral correlation length and small roughness \label{subsec:SmallLateralSmallRoughness}}

\begin{figure*}
\includegraphics[scale=0.4]{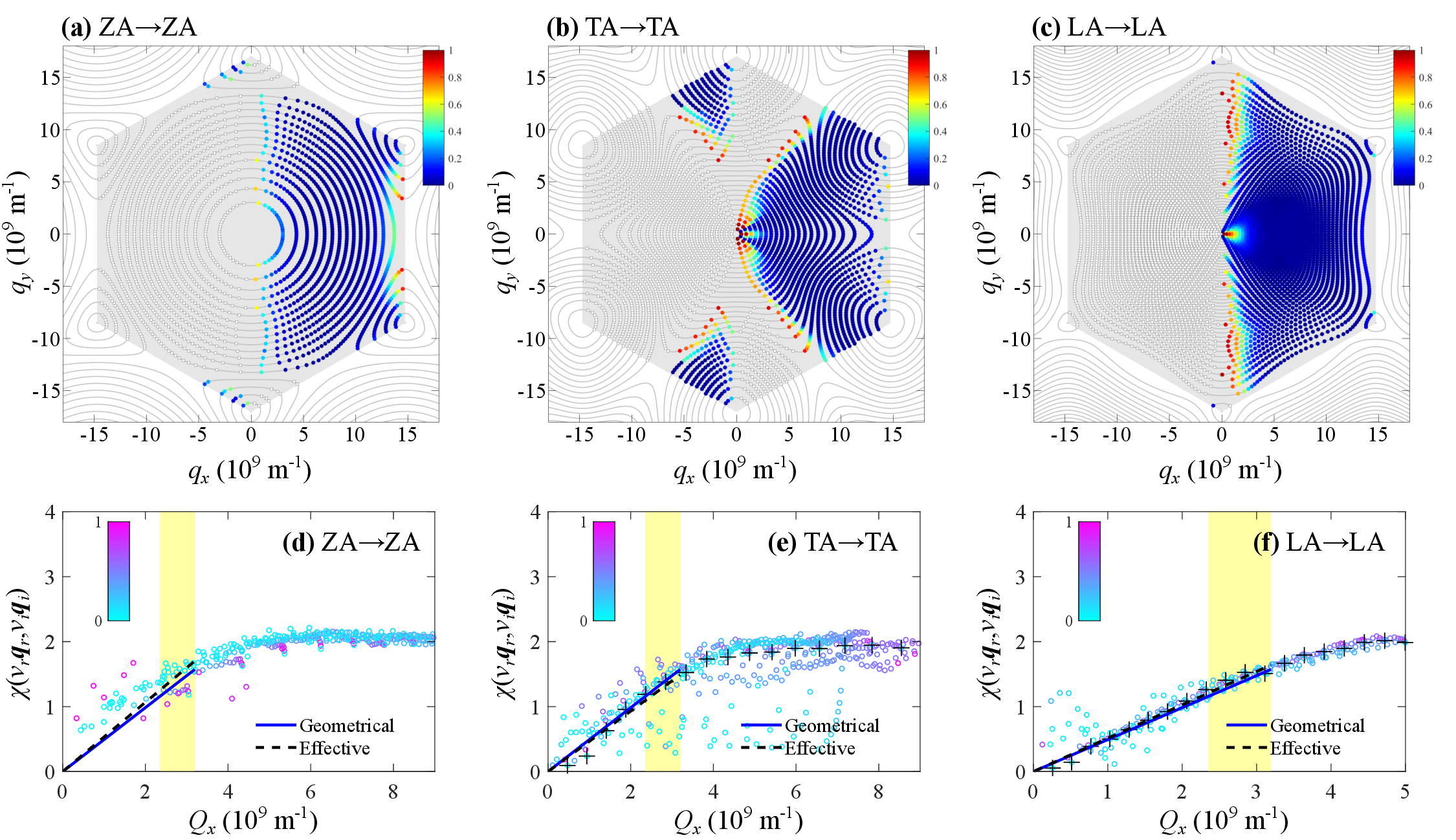}

\caption{Plot of the probability of specular reflectance $p_{\sigma}(\nu_{r}\boldsymbol{q}_{r},\nu_{i}\boldsymbol{q}_{i})$
with no mode conversion ($\nu_{r}=\nu_{i}$ and $|q_{r}^{\perp}|=|q_{i}^{\perp}|$)
for the (a) ZA, (b) TA and (c) LA phonons, distributed over the first
Brillouin zone, for $\sigma=0.5R_{0}$ and $L=L_{0}$ ($\mathcal{T}=\frac{1}{\sqrt{3}}$).
The value of each $p_{\sigma}(\nu_{r}\boldsymbol{q}_{r},\nu_{i}\boldsymbol{q}_{i})$
is indicated in color according to the color bar in the top right
corner of each panel. The corresponding data for $\chi(\nu_{r}\boldsymbol{q}_{r},\nu_{i}\boldsymbol{q}_{i})$
vs. $Q_{x}$ are shown as hollow circles in panels (d) to (f). The
corresponding angle of incidence $\theta_{i}$ of each $\chi$ data
point (normalized by $\pi/2$) is indicated by color according to
the color bar in the top left corner of each panel, with $0$ and
$1$ corresponding to normal and grazing incidence. Two linear fits
are drawn for $0<Q_{x}<\frac{\pi}{4\sigma}$ : the ``Geometrical''
(solid blue line) and the ``Effective'' (dashed black line). The range
for $\frac{1}{L}<Q_{x}<\frac{\pi}{4\sigma}$ is indicated by the yellow-shaded
region. For the TA and LA phonons, the data points for $\theta_{i}=0$
(normal incidence) are also indicated by the ``+'' symbol.}

\label{fig:SpecularLowRoughSmallCorrelationNoMode}
\end{figure*}

In Fig.~\ref{fig:SpecularLowRoughSmallCorrelationNoMode}, we plot
$p_{\sigma}(\nu_{r}\boldsymbol{q}_{r},\nu_{i}\boldsymbol{q}_{i})$
with \emph{no mode conversion} ($\nu_{r}=\nu_{i}$ and $|q_{r}^{\perp}|=|q_{i}^{\perp}|$)
for the (a) ZA, (b) TA and (c) LA phonons at a small correlation length
($L=L_{0}$) and small boundary roughness ($\sigma=0.5R_{0}$). Like
in Fig.~\ref{fig:BaseReflectanceNoMode}, the $p_{\sigma}$ data
in Fig.~\ref{fig:SpecularLowRoughSmallCorrelationNoMode}(a) to (c)
correspond to the $\text{ZA}\rightarrow\text{ZA}$, $\text{TA}\rightarrow\text{TA}$
and $\text{LA}\rightarrow\text{LA}$ scattering processes, respectively.
The topographic parameter for this boundary structure is $\mathcal{T}=\frac{1}{\sqrt{3}}$
which suggests that the Ogilvy formula should be valid. By comparing
the $p_{\sigma}$ data in Fig.~\ref{fig:SpecularLowRoughSmallCorrelationNoMode}(a)
to (c) to the $p_{0}$ data in Fig.~\ref{fig:BaseReflectanceNoMode}(a)
to (c), we can observe the effects of boundary roughness scattering
on specular reflection ($q_{r}^{\parallel}=q_{i}^{\parallel}$). As
expected, the values of $p_{\sigma}(\nu_{r}\boldsymbol{q}_{r},\nu_{i}\boldsymbol{q}_{i})$
are markedly attenuated by boundary roughness scattering, in contrast
to the results for $p_{0}$ in Fig.~\ref{fig:BaseReflectanceNoMode}(a)
to (c). To analyze this attenuation in the specular reflection, we
also plot the corresponding values of $\chi(\nu_{r}\boldsymbol{q}_{r},\nu_{i}\boldsymbol{q}_{i})$
from Eq.~(\ref{eq:AttenuationParameter}) as a function of $Q_{x}$
from Eq.~(\ref{eq:DefinitionQxVariable}) in Fig.~\ref{fig:SpecularLowRoughSmallCorrelationNoMode}(d)
to (f). In the absence of mode conversion, we have $Q_{x}=|q_{i}^{\perp}|=q_{x}$. 

To compare the $\chi(\nu_{r}\boldsymbol{q}_{r},\nu_{i}\boldsymbol{q}_{i})$
data and the Ogilvy formula from Eq.~(\ref{eq:ProbOgilvyFormula}),
we also draw in Fig.~\ref{fig:SpecularLowRoughSmallCorrelationNoMode}(d)
to (f) two lines going from the origin to $Q_{x}=\frac{\pi}{4\sigma}$,
which is the maximum value of $Q_{x}$ at which Eq.~(\ref{eq:ProbOgilvyFormula})
is expected to hold as implied by the Rayleigh roughness criterion~\citep{JDeSanto:Book92_Scalar}.
The first line, which is based on the RMS boundary roughness $\sigma$
from Eq.~(\ref{eq:GaussianRoughness}) used to construct the atomistic
graphene boundary like in Fig.~\ref{fig:RoughGrapheneBoundary},
is labeled `Geometrical' and given by $2\sigma Q_{x}$. The second
line (`Effective') is a linear fit of the $\chi$ data between $0<Q_{x}<\frac{\pi}{4\sigma}$
through the origin, from which we extract the parameter 
\begin{equation}
\rho_{\text{fit}}=\frac{\chi(\nu_{r}\boldsymbol{q}_{r},\nu_{i}\boldsymbol{q}_{i})}{2Q_{x}}\ .\label{eq:RoughnessFitParameter}
\end{equation}
Equation~(\ref{eq:RoughnessFitParameter}) describes the \emph{effective}
boundary roughness associated with phonon scattering as described
by Eq.~(\ref{eq:SpecularityAttenuationRatio}). We compare $\sigma$
and $\rho_{\text{fit}}$ to analyze the degree of agreement between
the geometrical boundary roughness $\sigma$, which describes scalar
wave scattering, and the effective boundary roughness $\rho_{\text{fit}}$
deduced from the $\chi$ data. Strictly speaking, the Ogilvy formula
is only valid in the $\frac{1}{L}<Q_{x}<\frac{\pi}{4\sigma}$ range,
which we indicate in the yellow-shaded region in Fig.~\ref{fig:SpecularLowRoughSmallCorrelationNoMode}(d)
to (f). 

In Fig.~\ref{fig:SpecularLowRoughSmallCorrelationNoMode}(d) to (f),
we observe a close agreement between $\sigma$ and $\rho_{\text{fit}}$.
When $Q_{x}<\frac{\pi}{4\sigma}$ and especially in the narrower $\frac{1}{L}<Q_{x}<\frac{\pi}{4\sigma}$
range, the Ogilvy formula describes the behavior of $\chi(\nu_{r}\boldsymbol{q}_{r},\nu_{i}\boldsymbol{q}_{i})$
which increases linearly with $Q_{x}$. This implies that the Ogilvy
formula provides a good description of the attenuation of the specularity
parameter for ZA, TA and LA phonons if there is no mode conversion.
For $Q_{x}>\frac{\pi}{4\sigma}$, the Rayleigh roughness criterion~\citep{JDeSanto:Book92_Scalar}
is no longer satisfied and the derivative of $\chi(\nu_{r}\boldsymbol{q}_{r},\nu_{i}\boldsymbol{q}_{i})$
decreases substantially, with the value of $\chi(\nu_{r}\boldsymbol{q}_{r},\nu_{i}\boldsymbol{q}_{i})$
plateauing at higher values of $Q_{x}$. This plateauing indicates
that the boundary roughness-induced attenuation of the specularity
parameter is maximized at a limiting value of $Q_{x}$. 

Nonetheless, there is a wide dispersion of the $\chi(\nu_{r}\boldsymbol{q}_{r},\nu_{i}\boldsymbol{q}_{i})$
data points around the $\sigma$ and $\rho_{\text{fit}}$ lines, especially
for the TA phonons when $Q_{x}$ is small. At small $Q_{x}$ and especially
for the LA phonon, the value of $\chi(\nu_{r}\boldsymbol{q}_{r},\nu_{i}\boldsymbol{q}_{i})$
is noticeably higher than that predicted by the Ogilvy formula. This
indicates that the attenuation of the specularity is greater than
what the Ogilvy formula predicts for these small-$Q_{x}$ (long-wavelength)
phonon modes and implies that the validity of Eq.~(\ref{eq:ProbOgilvyFormula})
for describing the specularity attenuation is limited when $L$ is
small and outside the $\frac{1}{L}<Q_{x}<\frac{\pi}{4\sigma}$ range.
This is probably due to the $qL\gg1$ condition associated with the
Kirchhoff Approximation which implies that the Ogilvy formula is only
valid when the lateral correlation length is much larger than the
wavelength.

The dispersion is more pronounced for the TA and LA phonons in Fig.~\ref{fig:SpecularLowRoughSmallCorrelationNoMode}(e)
and (f) but less so for the ZA phonons in Fig.~\ref{fig:SpecularLowRoughSmallCorrelationNoMode}(d).
In particular for the small-$Q_{x}$ LA phonons, the $\chi(\nu_{r}\boldsymbol{q}_{r},\nu_{i}\boldsymbol{q}_{i})$
data points can be significantly larger than $2\sigma Q_{x}$, indicating
that the attenuation is greater than what is predicted by the Ogilvy
formula. The more pronounced dispersion of the $\chi$ data could
be due to the effects of mode conversion ($\text{TA}\rightarrow\text{LA}$
and $\text{LA}\rightarrow\text{TA}$) which is absent for the boundary
roughness scattering of the ZA phonon. To analyze this, we also replot
the $\chi(\nu_{r}\boldsymbol{q}_{r},\nu_{i}\boldsymbol{q}_{i})$ data
for $\theta_{i}=0$, which corresponds to phonons at normal incidence
to the boundary, in Fig.~\ref{fig:SpecularLowRoughSmallCorrelationNoMode}(e)
and (f), and we find that they fall very close to the lines for $\sigma$
and $\rho_{\text{fit}}$. The plateauing of $\chi(\nu_{r}\boldsymbol{q}_{r},\nu_{i}\boldsymbol{q}_{i})$
at larger $Q_{x}$ values is also less ambiguous. This suggests that
the Ogilvy formula is more accurate for describing the boundary roughness
scattering of phonons at normal incidence to the boundary. 

\begin{figure*}
\includegraphics[scale=0.4]{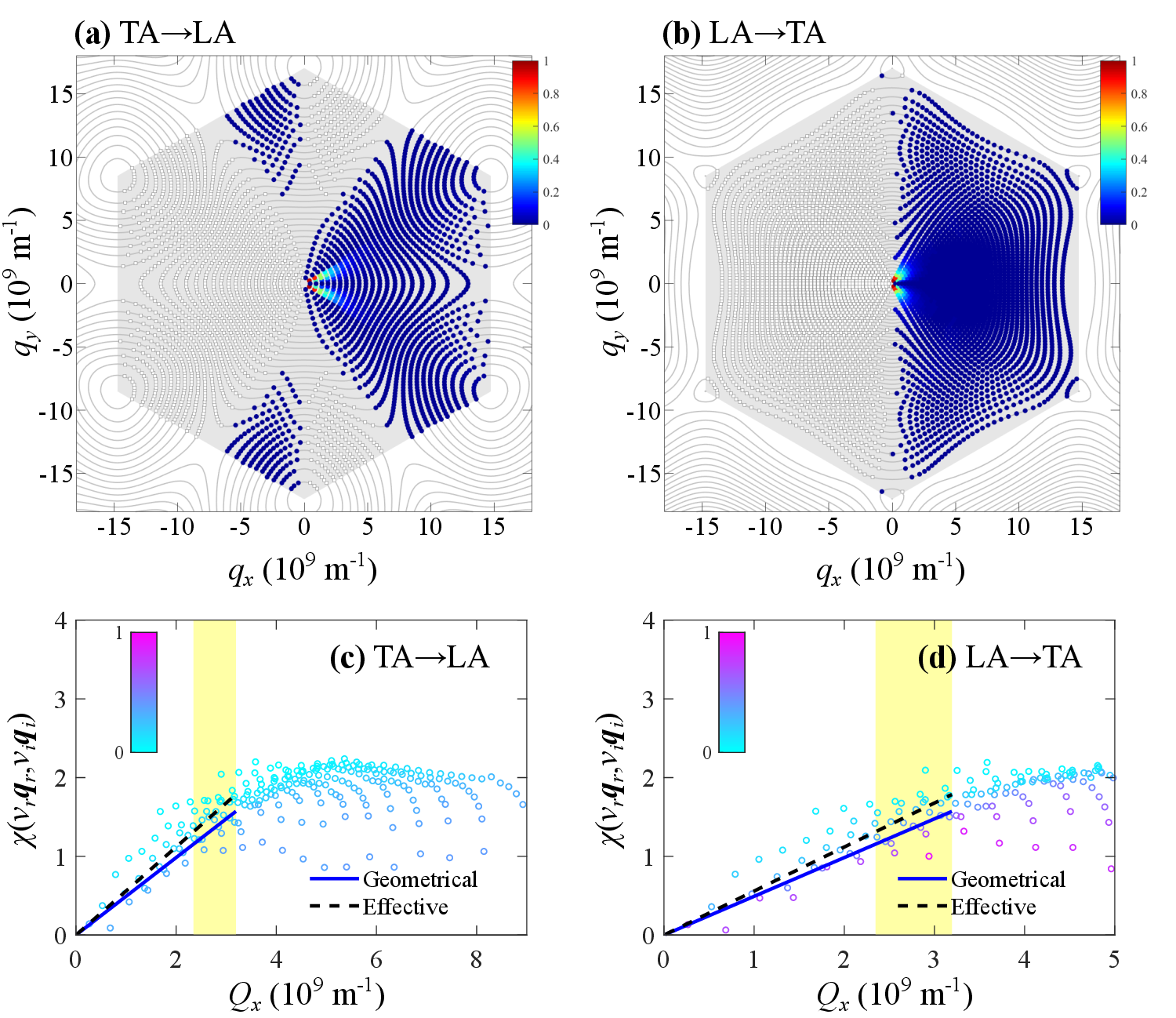}

\caption{Plot of the probability of specular reflectance $p_{\sigma}(\nu_{r}\boldsymbol{q}_{r},\nu_{i}\boldsymbol{q}_{i})$
with mode conversion ($\nu_{r}\protect\neq\nu_{i}$ and $|q_{r}^{\perp}|\protect\neq|q_{i}^{\perp}|$)
for incident (a) TA and (b) LA phonons, distributed over the first
Brillouin zone, for $\sigma=0.5R_{0}$ and $L=L_{0}$ ($\mathcal{T}=\frac{1}{\sqrt{3}}$).
The value of each $p_{\sigma}(\nu_{r}\boldsymbol{q}_{r},\nu_{i}\boldsymbol{q}_{i})$
is indicated in color according to the color bar in the top right
corner of each panel. The corresponding data for $\chi(\nu_{r}\boldsymbol{q}_{r},\nu_{i}\boldsymbol{q}_{i})$
are shown as hollow circles in panels (c) and (d). The corresponding
angle of incidence $\theta_{i}$ of each $\chi$ data point (normalized
by $\pi/2$) is indicated by color according to the color bar in the
top left corner of each panel.}

\label{fig:SpecularLowRoughSmallCorrelationModeConv}
\end{figure*}

For a more complete picture of the validity of the Ogilvy formula
for this boundary structure, we also study the attenuation of $p_{\sigma}(\nu_{r}\boldsymbol{q}_{r},\nu_{i}\boldsymbol{q}_{i})$
during \emph{mode conversion} ($\nu_{r}\neq\nu_{i}$ and $|q_{r}^{\perp}|\neq|q_{i}^{\perp}|$)
in which the polarization of the incident phonon is changed by boundary
scattering. Figure~\ref{fig:SpecularLowRoughSmallCorrelationModeConv}(a)
and (b) show the $p_{\sigma}(\nu_{r}\boldsymbol{q}_{r},\nu_{i}\boldsymbol{q}_{i})$
data with \emph{mode conversion} for the (a) TA and (b) LA phonons.
The $p_{\sigma}$ data in Fig.~\ref{fig:SpecularLowRoughSmallCorrelationModeConv}(a)
and (b) correspond to the $\text{TA}\rightarrow\text{LA}$ and $\text{LA}\rightarrow\text{TA}$
scattering processes, respectively. The range of incident angles for
the TA and LA phonons is given by $0<\theta_{i}<\theta_{\text{c}}$
and $0<\theta_{i}<\pi/2$, respectively. We also plot the corresponding
$\chi(\nu_{r}\boldsymbol{q}_{r},\nu_{i}\boldsymbol{q}_{i})$ data
as a function of $Q_{x}$ in Fig.~\ref{fig:SpecularLowRoughSmallCorrelationModeConv}(c)
and (d). 

In Fig.~\ref{fig:SpecularLowRoughSmallCorrelationModeConv}, we observe
that $\chi(\nu_{r}\boldsymbol{q}_{r},\nu_{i}\boldsymbol{q}_{i})$
increases linearly with $Q_{x}$ for $Q_{x}<\frac{\pi}{4\sigma}$
and plateaus at higher values of $Q_{x}$ like in Fig.~\ref{fig:SpecularLowRoughSmallCorrelationNoMode}(e)
and (f) although there is also a greater dispersion of the data points.
We observe that for $Q_{x}>\frac{\pi}{4\sigma}$, the attenuation
of $p_{\sigma}$ is weaker as the angle of incidence $\theta_{i}$
increases. We also find that the extracted value for the effective
boundary roughness $\rho_{\text{fit}}$ is also close but slighter
larger than the geometrical boundary roughness $\sigma$ for both
the TA and LA phonons. Compared to Fig.~\ref{fig:SpecularLowRoughSmallCorrelationNoMode}(e)
and (f), there is a smaller dispersion of the $\chi(\nu_{r}\boldsymbol{q}_{r},\nu_{i}\boldsymbol{q}_{i})$
data points around the $\rho_{\text{fit}}$ and $\sigma$ lines. This
suggests that the Ogilvy formula provides a better description of
the specularity attenuation from boundary roughness scattering for
scattering processes that involve mode conversion (e.g. $\text{TA}\rightarrow\text{LA}$
and $\text{LA}\rightarrow\text{TA}$). In addition, we notice a noticeable
dependence of $\chi$ on the angle of incidence $\theta_{i}$ especially
when $Q_{x}>\frac{\pi}{4\sigma}$. For the same $Q_{x}$ but higher
$\theta_{i}$ (or more oblique angle), $\chi(\nu_{r}\boldsymbol{q}_{r},\nu_{i}\boldsymbol{q}_{i})$
is smaller even though the Ogilvy formula from Eq.~(\ref{eq:ProbOgilvyFormula})
implies that it should only depend on $Q_{x}$.

\subsubsection{Small lateral correlation length and large roughness \label{subsec:SmallLateralLargeRoughness}}

\begin{figure*}
\includegraphics[scale=0.4]{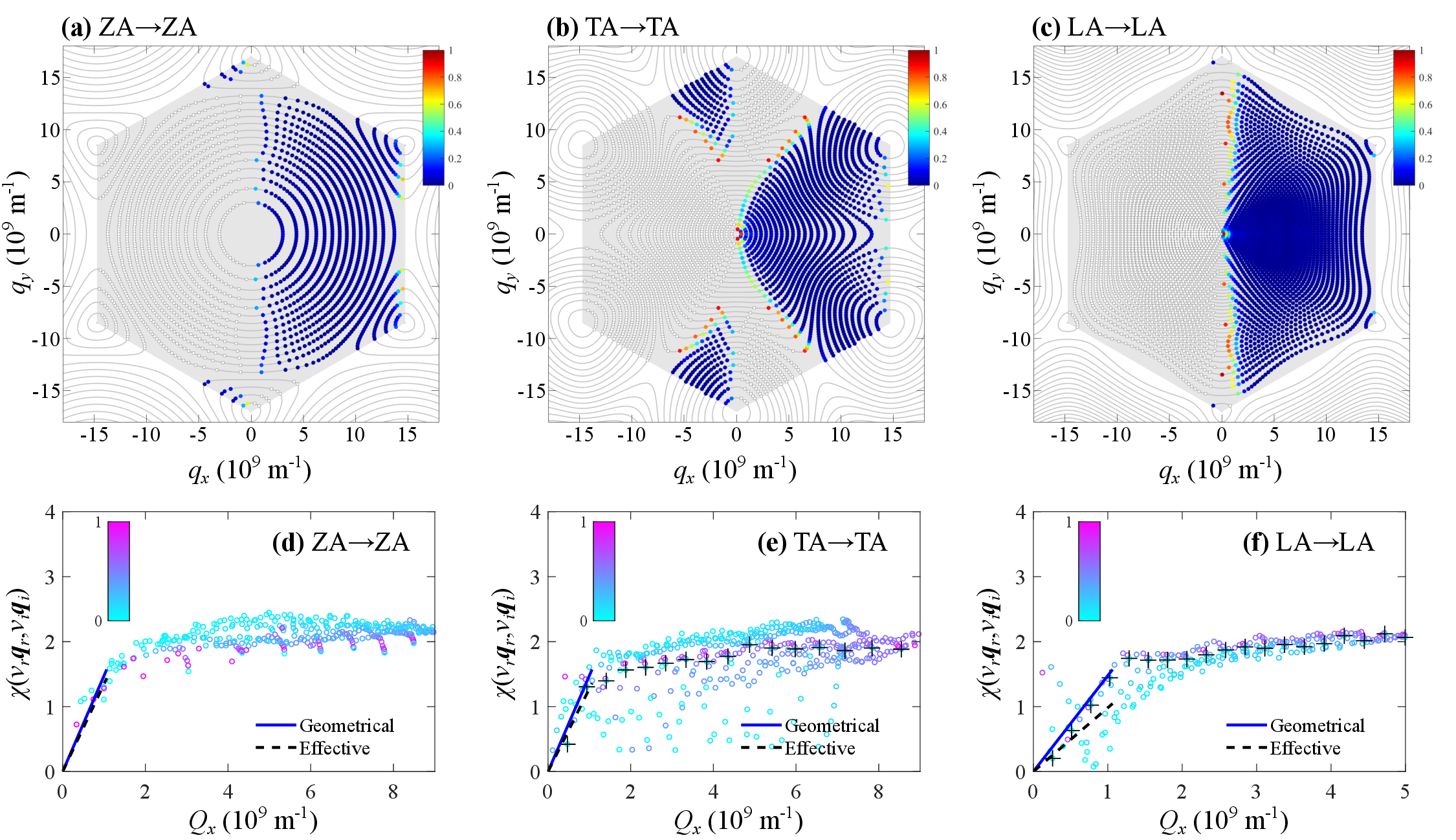}

\caption{Plot of the probability of specular reflectance $p_{\sigma}(\nu_{r}\boldsymbol{q}_{r},\nu_{i}\boldsymbol{q}_{i})$
with no mode conversion ($\nu_{r}=\nu_{i}$ and $|q_{r}^{\perp}|=|q_{i}^{\perp}|$)
for incident (a) ZA, (b) TA and (c) LA phonons, distributed over the
first Brillouin zone, for $\sigma=1.5R_{0}$ and $L=L_{0}$ ($\mathcal{T}=\sqrt{3}$).
The value of each $p_{\sigma}(\nu_{r}\boldsymbol{q}_{r},\nu_{i}\boldsymbol{q}_{i})$
is indicated in color according to the color bar in the top right
corner of each panel. The corresponding data for $\chi(\nu_{r}\boldsymbol{q}_{r},\nu_{i}\boldsymbol{q}_{i})$
are shown as hollow circles in panels (d) to (f). The corresponding
angle of incidence $\theta_{i}$ of each $\chi$ data point (normalized
by $\pi/2$) is indicated by color according to the color bar in the
top left corner of each panel. Two linear fits are drawn for $0<Q_{x}<\frac{\pi}{4\sigma}$
: the ``Geometrical'' (solid blue line) and the ``Effective'' (dashed
black line). For the TA and LA phonons, the data points for $\theta_{i}=0$
(normal incidence) are also indicated by the ``+'' symbol.}

\label{fig:SpecularHighRoughSmallCorrelationNoMode}
\end{figure*}

To see the effect of a larger boundary roughness, we repeat our analysis
of $p_{\sigma}(\nu_{r}\boldsymbol{q}_{r},\nu_{i}\boldsymbol{q}_{i})$
with \emph{no mode conversion} for the (a) ZA, (b) TA and (c) LA phonons
at the same small correlation length ($L=L_{0}$) as in Sec.~\ref{subsec:SmallLateralSmallRoughness}
but a larger boundary roughness ($\sigma=1.5R_{0}$). The results
are shown in Fig.~\ref{fig:SpecularHighRoughSmallCorrelationNoMode}.
The topographic parameter for this boundary structure is $\mathcal{T}=\sqrt{3}$,
which suggests that the Ogilvy formula in Eq.~(\ref{eq:ProbOgilvyFormula})
should \emph{not} be valid at all. Nevertheless, it would be interesting
to compare the degree of agreement between the corresponding $\chi(\nu_{r}\boldsymbol{q}_{r},\nu_{i}\boldsymbol{q}_{i})$
data and the Ogilvy formula. The data for $\chi(\nu_{r}\boldsymbol{q}_{r},\nu_{i}\boldsymbol{q}_{i})$
are shown in Figs.~\ref{fig:SpecularHighRoughSmallCorrelationNoMode}(d)
to \ref{fig:SpecularHighRoughSmallCorrelationNoMode}(f). Although
we have $\mathcal{T}=\sqrt{3}$, there is nonetheless some qualitative
agreement between the Ogilvy formula and the data. We observe that
$\chi(\nu_{r}\boldsymbol{q}_{r},\nu_{i}\boldsymbol{q}_{i})$ increases
with $Q_{x}$ up to $Q_{x}=\frac{\pi}{4\sigma}$ although there is
greater dispersion of the $\chi(\nu_{r}\boldsymbol{q}_{r},\nu_{i}\boldsymbol{q}_{i})$
data points around the lines for $\sigma$ and $\rho_{\text{fit}}$
which we can attribute to the greater $\sigma$. 

In Fig.~\ref{fig:SpecularHighRoughSmallCorrelationNoMode}(f) which
corresponds to LA phonons, the lines for $\sigma$ and $\rho_{\text{fit}}$
are markedly different with the effective roughness $\rho_{\text{fit}}$
being significantly smaller than the geometrical roughness $\sigma$.
In particular, we observe that at small $Q_{x}$, the $\chi(\nu_{r}\boldsymbol{q}_{r},\nu_{i}\boldsymbol{q}_{i})$
data for smaller $\theta_{i}$ but not at normal incidence tend to
be significantly lower than the values predicted by the Ogilvy formula.
In other words, specular reflection of the LA phonons is less attenuated
than predicted for slightly oblique angles. The $\chi(\nu_{r}\boldsymbol{q}_{r},\nu_{i}\boldsymbol{q}_{i})$
data for normal incidence or $\theta_{i}=0$ however show a much closer
fit to the line for $\sigma$. This suggests that the Ogilvy formula
in Eq.~(\ref{eq:ProbOgilvyFormula}) retains some accuracy for describing
the effects of boundary roughness scattering on the specularity, especially
for phonons at normal incidence, even when the boundary roughness
and topographic parameter are large enough to invalidate the Ogilvy
formula.

\begin{figure*}
\includegraphics[scale=0.4]{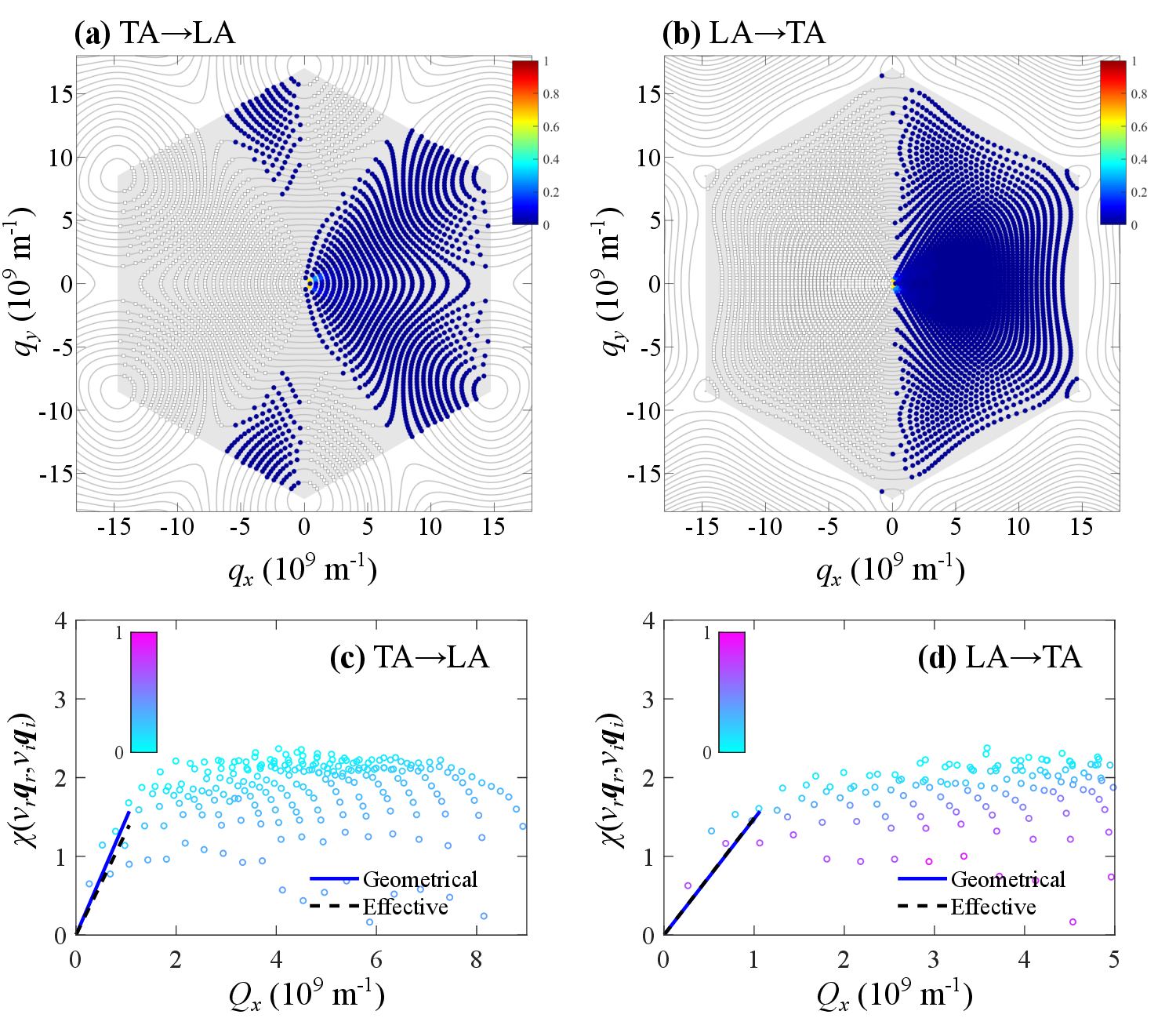}

\caption{Plot of the probability of specular reflectance $p_{\sigma}(\nu_{r}\boldsymbol{q}_{r},\nu_{i}\boldsymbol{q}_{i})$
with mode conversion ($\nu_{r}\protect\neq\nu_{i}$ and $|q_{r}^{\perp}|\protect\neq|q_{i}^{\perp}|$)
for the (a) TA and (b) LA phonons, distributed over the first Brillouin
zone, for $\sigma=1.5R_{0}$ and $L=L_{0}$ ($\mathcal{T}=\sqrt{3}$).
The value of each $p_{\sigma}(\nu_{r}\boldsymbol{q}_{r},\nu_{i}\boldsymbol{q}_{i})$
is indicated in color according to the color bar in the top right
corner of each panel. The corresponding data for $\chi(\nu_{r}\boldsymbol{q}_{r},\nu_{i}\boldsymbol{q}_{i})$
are shown as hollow circles in panels (c) and (d). The corresponding
angle of incidence $\theta_{i}$ of each $\chi$ data point (normalized
by $\pi/2$) is indicated by color according to the color bar in the
top left corner of each panel. }

\label{fig:SpecularHighRoughSmallCorrelationModeConv}
\end{figure*}

Figure~\ref{fig:SpecularHighRoughSmallCorrelationModeConv} shows
the data for $p_{\sigma}(\nu_{r}\boldsymbol{q}_{r},\nu_{i}\boldsymbol{q}_{i})$
with \emph{mode conversion} ($\nu_{r}\neq\nu_{i}$ and $|q_{r}^{\perp}|\neq|q_{i}^{\perp}|$)
for the (a) TA and (b) LA phonons for the same boundary structure.
Qualitatively, like in Fig.~\ref{fig:SpecularHighRoughSmallCorrelationNoMode},
we observe that $\chi(\nu_{r}\boldsymbol{q}_{r},\nu_{i}\boldsymbol{q}_{i})$
increases with $Q_{x}$ for $Q_{x}<\frac{\pi}{4\sigma}$. For $Q_{x}>\frac{\pi}{4\sigma}$,
we similarly observe that the maximum value of $\chi(\nu_{r}\boldsymbol{q}_{r},\nu_{i}\boldsymbol{q}_{i})$
plateaus with $Q_{x}$ although there are $\chi(\nu_{r}\boldsymbol{q}_{r},\nu_{i}\boldsymbol{q}_{i})$
data points in Figs.~\ref{fig:SpecularHighRoughSmallCorrelationModeConv}(c)
and \ref{fig:SpecularHighRoughSmallCorrelationModeConv}(d) that are
significantly below the plateau line. Like in Fig.~\ref{fig:SpecularLowRoughSmallCorrelationModeConv},
we similarly observe that for $Q_{x}>\frac{\pi}{4\sigma}$, the attenuation
of $p_{\sigma}$ is weaker as $\theta_{i}$ increases or becomes closer
to the grazing angle. This implies that $\chi(\nu_{r}\boldsymbol{q}_{r},\nu_{i}\boldsymbol{q}_{i})$
has a dependence on $\theta_{i}$ in addition to its dependence on
$Q_{x}$ when $\mathcal{T}$ is large, and that there is an additional
dependence on the transverse momentum $q_{y}$ that is not predicted
in the Ogilvy formula. By comparing Figs.~\ref{fig:SpecularLowRoughSmallCorrelationModeConv}
and \ref{fig:SpecularHighRoughSmallCorrelationModeConv}, we observe
that this angle-dependent weakening of the attenuation parameter is
more pronounced when $\sigma$ is larger. 

\subsubsection{Large lateral correlation length and small roughness \label{subsec:LargeLateralSmallRoughness}}

\begin{figure*}
\includegraphics[scale=0.4]{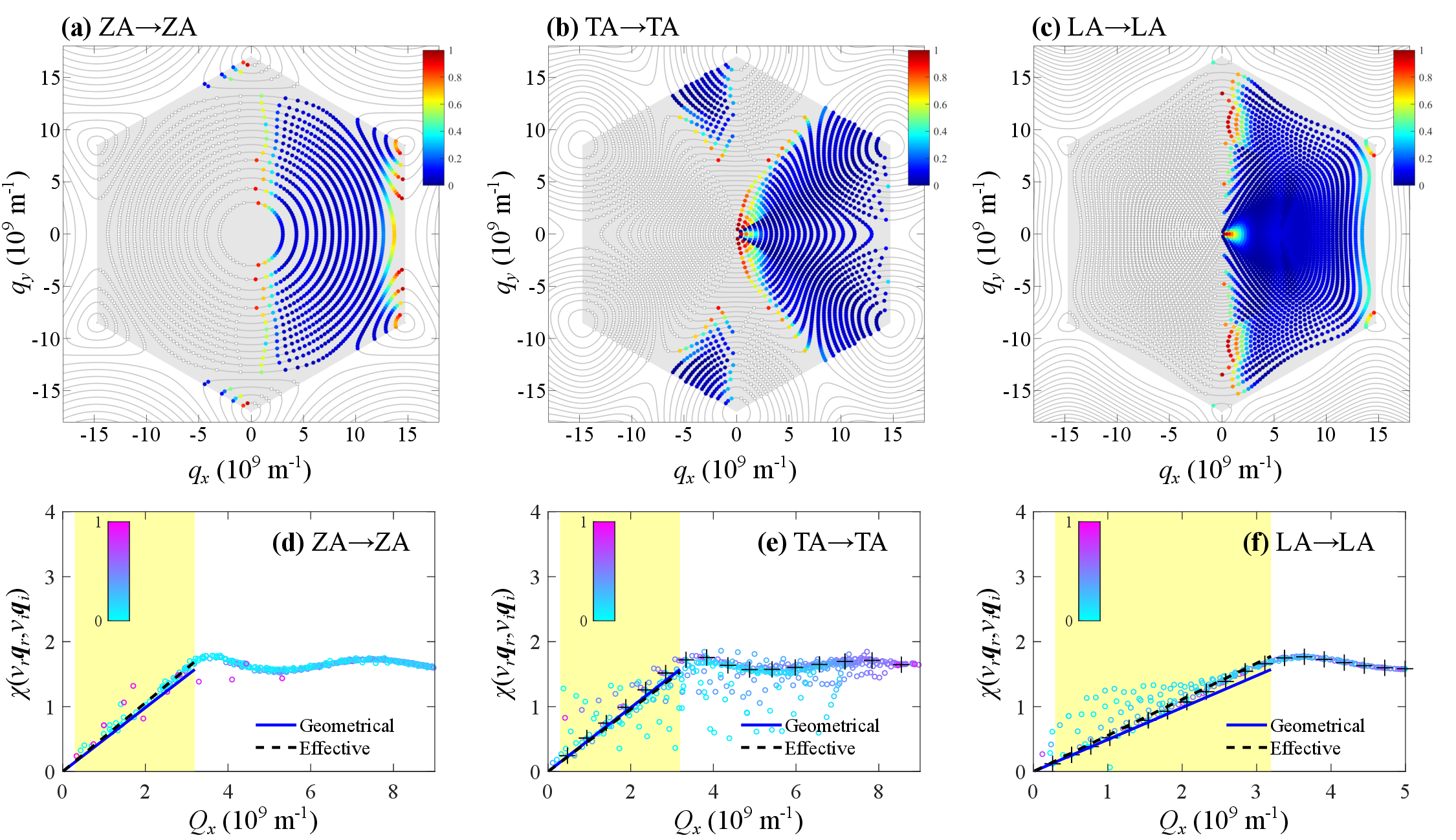}

\caption{Plot of the probability of specular reflectance $p_{\sigma}(\nu_{r}\boldsymbol{q}_{r},\nu_{i}\boldsymbol{q}_{i})$
with no mode conversion ($\nu_{r}=\nu_{i}$ and $|q_{r}^{\perp}|=|q_{i}^{\perp}|$)
for the (a) ZA, (b) TA and (c) LA phonons, distributed over the first
Brillouin zone, for $\sigma=0.5R_{0}$ and $L=8L_{0}$ ($\mathcal{T}=\frac{1}{8\sqrt{3}}$).
The value of each $p_{\sigma}(\nu_{r}\boldsymbol{q}_{r},\nu_{i}\boldsymbol{q}_{i})$
is indicated in color according to the color bar in the top right
corner of each panel. The corresponding data for $\chi(\nu_{r}\boldsymbol{q}_{r},\nu_{i}\boldsymbol{q}_{i})$
vs. $Q_{x}$ are shown as hollow circles in panels (d) to (f). The
corresponding angle of incidence $\theta_{i}$ of each $\chi$ data
point (normalized by $\pi/2$) is indicated by color according to
the color bar in the top left corner of each panel. Two linear fits
are drawn for $0<Q_{x}<\frac{\pi}{4\sigma}$ : the ``Geometrical''
(solid blue line) and the ``Effective'' (dashed black line). For the
TA and LA phonons, the data points for $\theta_{i}=0$ (normal incidence)
are also indicated by the ``+'' symbol.}

\label{fig:SpecularLowRoughLargeCorrelationNoMode}
\end{figure*}

At large lateral correlation lengths, the boundary is smoother when
$L\gg\sigma$ and $\mathcal{T}\ll1$. In Fig.~\ref{fig:SpecularLowRoughLargeCorrelationNoMode},
we plot $p_{\sigma}(\nu_{r}\boldsymbol{q}_{r},\nu_{i}\boldsymbol{q}_{i})$
with \emph{no mode conversion} for the (a) ZA, (b) TA and (c) LA phonons
at a large correlation length ($L=8L_{0}$) and a small boundary roughness
($\sigma=0.5R_{0}$) as in Sec.~\ref{subsec:SmallLateralSmallRoughness}.
We also plot the corresponding values of $\chi(\nu_{r}\boldsymbol{q}_{r},\nu_{i}\boldsymbol{q}_{i})$
from Eq.~(\ref{eq:SpecularityAttenuationRatio})in Figs.~\ref{fig:SpecularLowRoughLargeCorrelationNoMode}(d)
to \ref{fig:SpecularLowRoughLargeCorrelationNoMode}(f). Because the
corresponding topographic parameter $\mathcal{T}=\frac{1}{8\sqrt{3}}\ll1$
is much smaller than unity, we expect Eq.~(\ref{eq:ProbOgilvyFormula})
to be optimal for describing specularity attenuation by boundary roughness
scattering. Thus, by comparing the $\chi(\nu_{r}\boldsymbol{q}_{r},\nu_{i}\boldsymbol{q}_{i})$
data and the Ogilvy formula, we can assess the accuracy of Eq.~(\ref{eq:ProbOgilvyFormula})
in the most ideal case.

Compared to the results for a small correlation length ($L=L_{0}$)
and small boundary roughness ($\sigma=0.5R_{0}$) in Figs.~\ref{fig:SpecularLowRoughSmallCorrelationNoMode}(d)
to \ref{fig:SpecularLowRoughSmallCorrelationNoMode}(f) where $\mathcal{T}=\frac{1}{\sqrt{3}}$,
there is a much smaller dispersion of the $\chi(\nu_{r}\boldsymbol{q}_{r},\nu_{i}\boldsymbol{q}_{i})$
data points around the line for the geometrical boundary roughness
$\sigma$, especially for the ZA phonons where most of the $\chi$
data points tend to fall very close to the line when $Q_{x}<\frac{\pi}{4\sigma}$.
In fact, one observes a near-perfect agreement with Eq.~(\ref{eq:ProbOgilvyFormula})
for the ZA phonons when $Q_{x}<\frac{\pi}{4\sigma}$ even though the
ZA phonons have a nonlinear phonon dispersion, with $\omega\propto q^{2}$
in the long wavelength limit. Nonetheless, we still observe that at
small $Q_{x}$ for LA phonons in Fig.~\ref{fig:SpecularLowRoughLargeCorrelationNoMode}(f),
the value of $\chi(\nu_{r}\boldsymbol{q}_{r},\nu_{i}\boldsymbol{q}_{i})$
is noticeably higher than that predicted by the Ogilvy formula, i.e.,
the specularity of the small-$Q_{x}$ modes is more strongly attenuated
than what is predicted by the Ogilvy formula. This is similar to what
we observe in Fig.~\ref{fig:SpecularLowRoughSmallCorrelationNoMode}(f)
for a smaller lateral correlation length ($L=L_{0}$). 

The geometrical and effective boundary roughness values $\sigma$
and $\rho_{\text{fit}}$ are also in good agreement. Qualitatively,
Figs.~\ref{fig:SpecularLowRoughLargeCorrelationNoMode}(d) to \ref{fig:SpecularLowRoughLargeCorrelationNoMode}(f)
suggest that the agreement with the Ogilvy formula in Eq.~(\ref{eq:ProbOgilvyFormula})
is greater when the lateral correlation length $L$ is large. This
is expected since Eq.~(\ref{eq:ProbOgilvyFormula}) is derived assuming
that the correlation length is much greater than the wavelength or
$qL\gg1$. In addition, Figs.~\ref{fig:SpecularLowRoughLargeCorrelationNoMode}(e)
and \ref{fig:SpecularLowRoughLargeCorrelationNoMode}(f) show that
for TA and LA phonons, the $\chi(\nu_{r}\boldsymbol{q}_{r},\nu_{i}\boldsymbol{q}_{i})$
data points for $\theta_{i}=0$ also have excellent agreement with
Eq.~(\ref{eq:ProbOgilvyFormula}). 

Beyond the $Q_{x}=\frac{\pi}{4\sigma}$ point, we observe a clear
plateauing of $\chi(\nu_{r}\boldsymbol{q}_{r},\nu_{i}\boldsymbol{q}_{i})$
at a maximum value of $\chi\sim\frac{\pi}{2}$, similar to that in
Figs.~\ref{fig:SpecularLowRoughSmallCorrelationNoMode}(d) to \ref{fig:SpecularLowRoughSmallCorrelationNoMode}(f)
where $L$ is smaller. This suggests that relative to $p_{0}$, there
possibly exists a minimum value for $p_{\sigma}$ associated with
the maximum attenuation of short-wavelength acoustic phonons, which
we can estimate as 
\begin{equation}
\lim_{q_{i}^{\perp}\rightarrow\infty}p_{\sigma}(\nu_{r}\boldsymbol{q}_{r},\nu_{i}\boldsymbol{q}_{i})\sim p_{0}(\nu_{r}\boldsymbol{q}_{r},\nu_{i}\boldsymbol{q}_{i})\exp(-\frac{\pi^{2}}{4})\ .\label{eq:MinimumSpecularity}
\end{equation}
If the minimum $p_{\sigma}$ in Eq.~(\ref{eq:MinimumSpecularity})
does exist, it leads to the question of whether this phenomenon is
peculiar to the boundary roughness scattering of short-wavelength
phonons or if it can be generalized to the boundary roughness scattering
of other quasiparticles or waves in general. In the context of phonon-mediated
thermal transport, this implies that a boundary scattering event cannot
completely dissipate all the momentum of the incident phonon. 

\begin{figure*}
\includegraphics[scale=0.4]{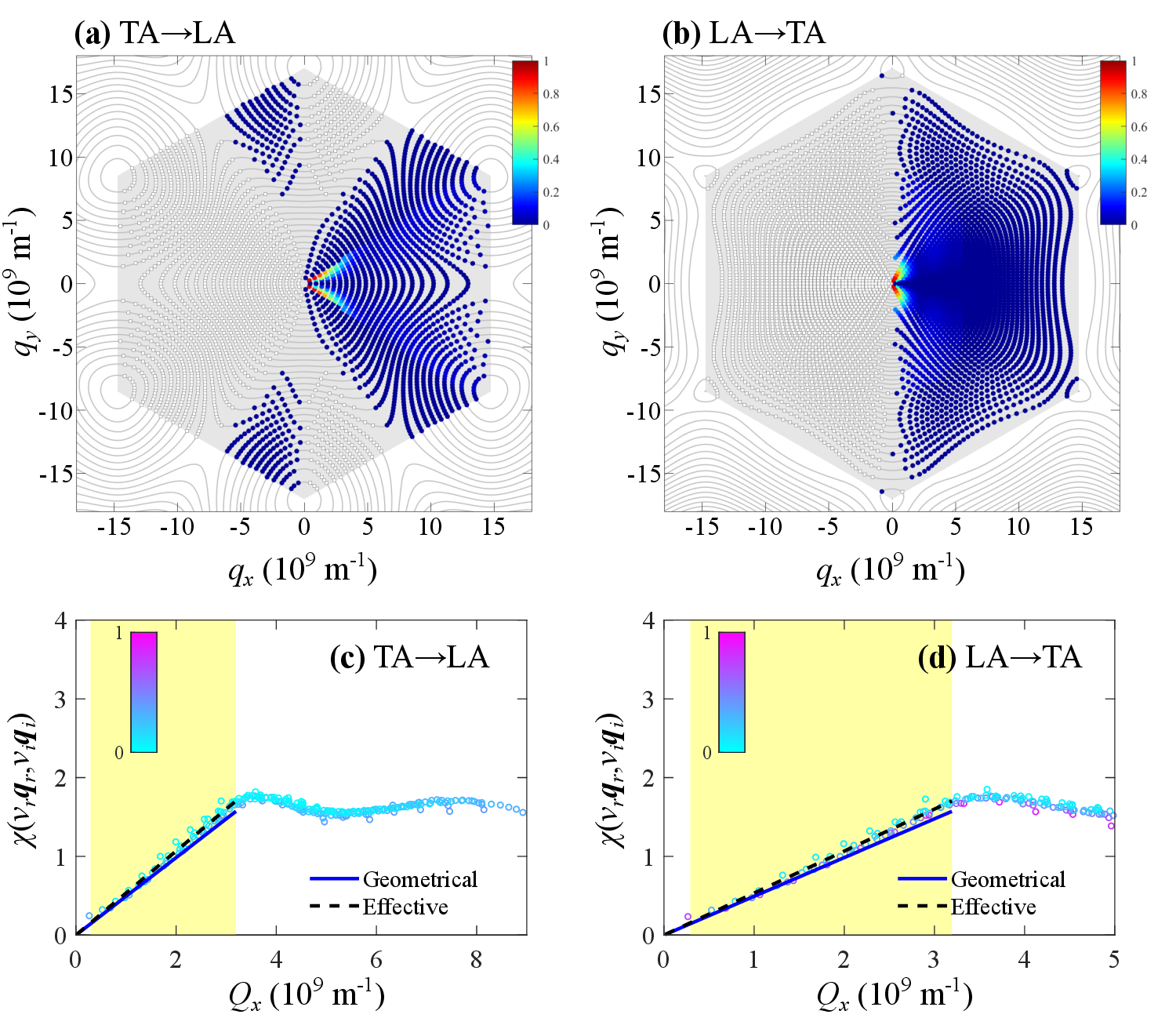}

\caption{Plot of the probability of specular reflectance $p_{\sigma}(\nu_{r}\boldsymbol{q}_{r},\nu_{i}\boldsymbol{q}_{i})$
with mode conversion ($\nu_{r}\protect\neq\nu_{i}$ and $|q_{r}^{\perp}|\protect\neq|q_{i}^{\perp}|$)
for the (a) TA and (b) LA phonons, distributed over the first Brillouin
zone, for $\sigma=0.5R_{0}$ and $L=8L_{0}$ ($\mathcal{T}=\frac{1}{8\sqrt{3}}$).
The value of each $p_{\sigma}(\nu_{r}\boldsymbol{q}_{r},\nu_{i}\boldsymbol{q}_{i})$
is indicated in color according to the color bar in the top right
corner of each panel. The corresponding data for $\chi(\nu_{r}\boldsymbol{q}_{r},\nu_{i}\boldsymbol{q}_{i})$
are shown as hollow circles in panels (c) to (d). The corresponding
angle of incidence $\theta_{i}$ of each $\chi$ data point (normalized
by $\pi/2$) is indicated by color according to the color bar in the
top left corner of each panel. }

\label{fig:SpecularLowRoughLargeCorrelationModeConv}
\end{figure*}

Figure~\ref{fig:SpecularLowRoughLargeCorrelationModeConv} shows
the corresponding data for $p_{\sigma}(\nu_{r}\boldsymbol{q}_{r},\nu_{i}\boldsymbol{q}_{i})$
with \emph{mode conversion} ($\nu_{r}\neq\nu_{i}$ and $|q_{r}^{\perp}|\neq|q_{i}^{\perp}|$)
for the (a) TA and (b) LA phonons in the same boundary structure.
We also plot $\chi(\nu_{r}\boldsymbol{q}_{r},\nu_{i}\boldsymbol{q}_{i})$
for (a) TA and (b) LA phonons. When $Q_{x}<\frac{\pi}{4\sigma}$,
the $\chi(\nu_{r}\boldsymbol{q}_{r},\nu_{i}\boldsymbol{q}_{i})$ data
points are in near-perfect agreement with Eq.~(\ref{eq:ProbOgilvyFormula}),
clustering close to the lines for $\sigma$ and $\rho_{\text{fit}}$.
This can be contrasted to the results in Figs.~\ref{fig:SpecularLowRoughLargeCorrelationNoMode}(e)
and \ref{fig:SpecularLowRoughLargeCorrelationNoMode}(f) for the $\text{TA}\rightarrow\text{TA}$
and $\text{LA}\rightarrow\text{LA}$ scattering processes, respectively.
We also observe the plateauing of the $\chi(\nu_{r}\boldsymbol{q}_{r},\nu_{i}\boldsymbol{q}_{i})$
data points when $Q_{x}>\frac{\pi}{4\sigma}$. Unlike the results
in Figs.~\ref{fig:SpecularLowRoughSmallCorrelationModeConv} and
\ref{fig:SpecularHighRoughSmallCorrelationModeConv} for $L=L_{0}$,
the attenuation of $p_{\sigma}$ does not vary significantly with
the angle of incidence $\theta_{i}$ for $Q_{x}>\frac{\pi}{4\sigma}$
in Fig.~\ref{fig:SpecularLowRoughLargeCorrelationModeConv}. This
suggests that the additional $\theta_{i}$-dependence of the $\chi(\nu_{r}\boldsymbol{q}_{r},\nu_{i}\boldsymbol{q}_{i})$
data seen in Figs.~\ref{fig:SpecularLowRoughSmallCorrelationModeConv}
and \ref{fig:SpecularHighRoughSmallCorrelationModeConv} becomes more
pronounced as the topographic parameter $\mathcal{T}$ increases. 

Given the small $\mathcal{T},$the results in Figs.~\ref{fig:SpecularLowRoughLargeCorrelationNoMode}
and \ref{fig:SpecularLowRoughLargeCorrelationModeConv} show that
the Ogilvy formula is generally excellent for describing the specularity
attenuation by boundary roughness scattering except when there is
significant possible mode conversion between the phonons involved
as with the TA and LA phonons. In the latter case, the formula appears
to work better to describe boundary roughness scattering with mode
conversion ($\text{TA}\rightarrow\text{LA}$ and $\text{LA}\rightarrow\text{TA}$)
than for scattering processes without mode conversion ($\text{LA}\rightarrow\text{LA}$
and $\text{TA}\rightarrow\text{TA}$). When there is no possible mode
conversion (e.g. $\text{ZA\ensuremath{\rightarrow\text{ZA}}}$ or
$\text{TA\ensuremath{\rightarrow\text{TA}}}$ and $\text{LA\ensuremath{\rightarrow\text{LA}}}$
at normal incidence), the Ogilvy formula is also more accurate. 

\section{Summary and conclusions}

For the convenience of the reader, we summarize our findings from
the simulation results from Sec. \ref{sec:SimulationResults}. In
our simulations, we distinguish and investigate two types of scattering
processes \textendash{} those without mode conversion ($\text{ZA}\rightarrow\text{ZA}$,
$\text{TA}\rightarrow\text{TA}$, and $\text{LA}\rightarrow\text{LA}$)
and those with mode conversion ($\text{TA}\rightarrow\text{LA}$,
and $\text{LA}\rightarrow\text{TA}$) \textendash{} for different
boundary structures characterized by the structural parameters $L$
and $\sigma$. The simulation results are benchmarked to the Ogilvy
formula from Eq.~(\ref{eq:ProbOgilvyFormula}), which describes the
attenuation of the specular reflection, and used to determine its
dependence on mode conversion and the boundary structural parameters.
A notable feature of the Ogilvy formula is that it depends only on
the longitudinal component of the incident and reflected wave vectors
as given in Eq.~(\ref{eq:DefinitionQxVariable}). 

In general, we find that the Ogilvy formula provides a reasonable
quantitative description of the attenuation parameter $\chi(\nu_{r}\boldsymbol{q}_{r},\nu_{i}\boldsymbol{q}_{i})$
from Eq.~(\ref{eq:AttenuationParameter}), with the data for $\chi$
exhibiting a linear dependence on $Q_{x}$ (i.e. $\chi\approx2\rho_{\text{fit}}Q_{x}$)
when $\frac{1}{L}<Q_{x}<\frac{\pi}{4\sigma}$. We find that the extracted
effective boundary roughness $\rho_{\text{fit}}$  is very close to
the geometrical boundary roughness $\sigma$ when $\mathcal{T}<1$.
This suggests that the Ogilvy formula applies to the boundary roughness
scattering of phonons. As expected, the degree of agreement between
the data for $\chi$ and the Ogilvy formula for $Q_{x}<\frac{\pi}{4\sigma}$
increases as the topographic parameter $\mathcal{T}$ decreases. Nonetheless,
even when $\mathcal{T}$ is large (i.e. $\mathcal{T}>1$), the Ogilvy
formula can still describe the linear dependence of $\chi(\nu_{r}\boldsymbol{q}_{r},\nu_{i}\boldsymbol{q}_{i})$
on $Q_{x}$ very well for phonon modes at normal incidence ($\theta_{i}=0$)
to the boundary. 

Beyond the $Q_{x}=\frac{\pi}{4\sigma}$ point, the value of $\chi$
plateaus at larger values of $Q_{x}$ with the asymptotic value of
$\chi$ associated with a \emph{possible} minimum $p_{\sigma}$ value
relative to $p_{0}$. This plateauing is more obvious when $\mathcal{T}\ll1$.
Hence, when $\mathcal{T}$ is small, our results suggest that there
are two observable regimes for $\chi$: a linear dependence on $Q_{x}$
or $\chi\propto Q_{x}$ for the small-$Q_{x}$ regime and an asymptotic
convergence to a constant $\chi\sim\frac{\pi}{2}$ for the large-$Q_{x}$
regime. We conjecture from our numerical results that there exists
an \emph{effective minimum specularity value} of $p_{\sigma}\sim p_{0}\exp(-\frac{\pi^{2}}{4})$
or a maximum attenuation $\chi\sim\frac{\pi}{2}$ for short-wavelength
acoustic phonons in the $\mathcal{T}\rightarrow0$ (smooth) limit.
This suggests that the momentum of the incident phonon cannot be totally
dissipated by boundary scattering. 

The effect of mode conversion presents two interesting phenomena.
Firstly, in scattering processes without mode conversion (Fig.~\ref{fig:SpecularLowRoughLargeCorrelationNoMode}),we
find that $\chi$ can be significantly larger than the predicted value
of $2\rho_{\text{fit}}Q_{x}$ at very small $Q_{x}$ for the LA and
TA phonons when $\mathcal{T}$ is small, i.e., the attenuation is
significantly stronger than what the Ogilvy formula predicts for long
wavelength phonons. Secondly, in scattering processes that involve
mode conversion ($\text{LA}\rightarrow\text{LA}$ and $\text{TA}\rightarrow\text{TA}$),
at large $Q_{x}$, $\chi$ can  vary independently with the angle
of incidence $\theta_{i}$ with $\chi$ decreasing as $\theta_{i}$
approaches the grazing angle of $\pi/2$ when $\mathcal{T}$ is large.
This implies that the attenuation of the specularity also depends
on $q_{y}$ the transverse component of the incident wave vector.
This dependence on $q_{y}$ also implies that $q_{y}$-dependent corrections
are needed for the Ogilvy formula when $\mathcal{T}$ is large.

In conclusion, our results shed light on the accuracy of the Ogilvy
formula in determining the extent that boundary roughness scattering
attenuates phonon specular reflection in graphene. They confirm that
it is generally accurate for the $Q_{x}<\frac{\pi}{4\sigma}$ regime
when $L$ is large. In the large-$Q_{x}$ (short wavelength) regime,
they suggest that the Ogilvy formula is not valid and that the attenuation
parameter $\chi(\nu_{r}\boldsymbol{q}_{r},\nu_{i}\boldsymbol{q}_{i})$
may exhibit a weak or no dependence on $Q_{x}$. 
\begin{acknowledgments}
We acknowledge funding support from the Agency for Science, Technology
and Research (A{*}STAR) of Singapore with the Manufacturing, Trade
and Connectivity (MTC) Programmatic Grant ``Advanced Modeling Models
for Additive Manufacturing'' (M22L2b0111) and from the Polymer Matrix
Composites Program (SERC Grant No. A19C9a004). We also acknowledge
the assistance of Wen Han Zhang from the National University of Singapore
in the preliminary phase of the study. We also acknowledge helpful
discussions on the Ogilvy formula with Fan Shi (Hong Kong University
of Science and Technology) and Stewart G. Haslinger (University of
Liverpool). 
\end{acknowledgments}

\bibliography{PaperReferences}

\end{document}